\documentclass[submission, Phys]{SciPost}
\usepackage{amsmath}
\usepackage{amssymb}
\newcommand{\vect}[1]{\boldsymbol{#1}}
\def\abs#1{\left|#1\right|}
\def\scaleFac{\Lambda}

\begin{document}

\begin{center}{\Large \textbf{
Magnetism of magic-angle twisted bilayer graphene
}}\end{center}

\begin{center}
Javad Vahedi\textsuperscript{1},
Robert Peters\textsuperscript{2},
Ahmed Missaoui\textsuperscript{3},
Andreas Honecker\textsuperscript{3*}, \\
Guy Trambly de Laissardi\`ere\textsuperscript{3}
\end{center}

\begin{center}
{\bf 1} Technische Universit\"at Braunschweig, Institut f\"ur Mathematische Physik, Mendelssohnstra{\ss}e 3,
 38106 Braunschweig, Germany
\\
{\bf 2} Department of Physics, Kyoto University, Kyoto 606-8502, Japan
\\
{\bf 3} Laboratoire de Physique Th\'eorique et Mod\'elisation, CNRS UMR 8089, \\
  CY Cergy Paris Universit\'e, 95302 Cergy-Pontoise Cedex, France
\\
* andreas.honecker@cyu.fr
\end{center}

\begin{center}
%\today
April 21, 2021; revised August 17, 2021
\end{center}

% For convenience during refereeing: line numbers
%\linenumbers

\section*{Abstract}
{\bf
We investigate magnetic instabilities in charge-neutral twisted bilayer 
graphene close to so-called ``magic angles'' using a combination of 
real-space Hartree-Fock and dynamical mean-field theories. In view of the 
large size of the unit cell close to magic angles, we examine a 
previously proposed rescaling that permits to mimic the same underlying 
flat minibands at larger twist angles. We find that localized magnetic 
states emerge for values of the Coulomb interaction $U$ that are 
significantly smaller than what would be required to render an isolated 
layer antiferromagnetic. However, this effect is overestimated in the 
rescaled system, hinting at a complex interplay of flatness of the 
minibands close to the Fermi level and the spatial extent of the 
corresponding localized states. Our findings shed new light on 
perspectives for experimental realization of magnetic states in 
charge-neutral twisted bilayer graphene.
}

\vspace{10pt}
\noindent\rule{\textwidth}{1pt}
\tableofcontents\thispagestyle{fancy}
\noindent\rule{\textwidth}{1pt}
\vspace{10pt}

\section{Introduction}
\label{sec:intro}

\begin{figure}[t!]
\centering
\qquad\quad\includegraphics[width=0.7\linewidth]{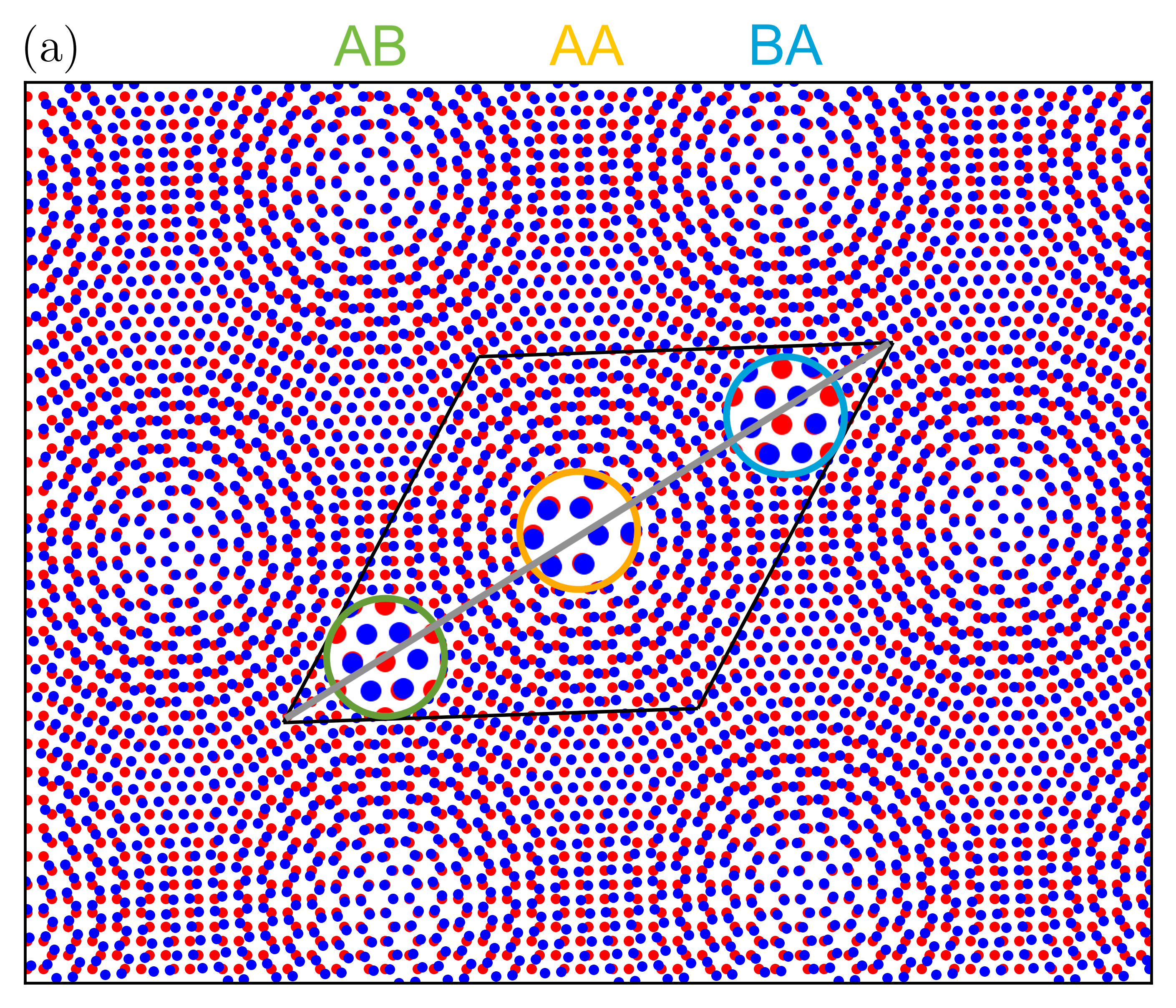}\\
\includegraphics[width=0.78\linewidth]{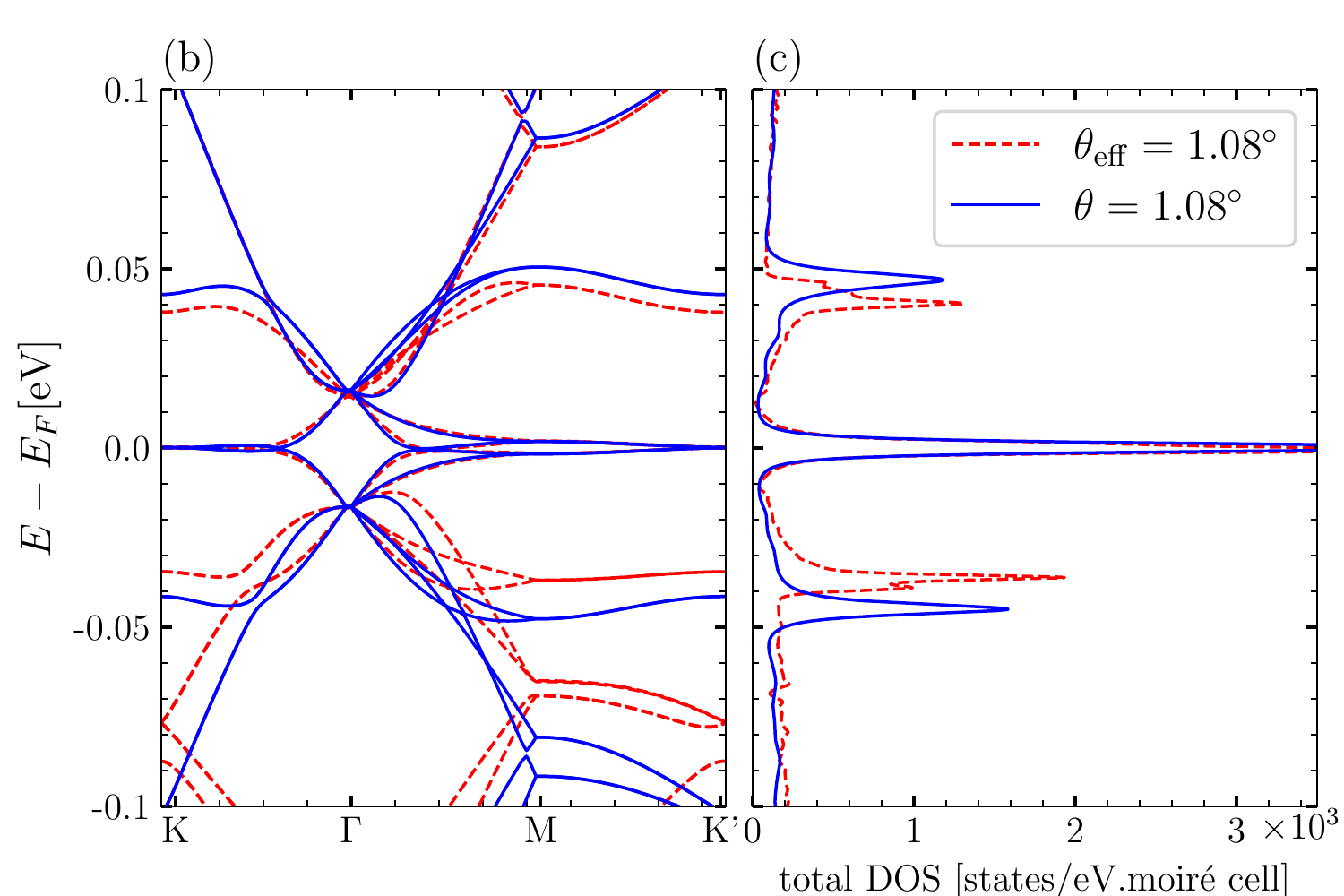}
\caption{(a) Moir\'e pattern for a twist angle $\theta=3.89^\circ$, 
[$(n,m)=(8,9)$] with the identification of magnified regions with AB, AA, and BA stacking.
(b) Band structure calculated for a system with 
$\theta=1.08^\circ$, [$(n,m)=(30,31)$] and $\theta_{\rm eff}=1.08^\circ$, 
[$(n,m)=(8,9)$], (c) total density of states (DOS) corresponding to panel (b). The almost 
flat minibands at zero energy and corresponding large DOS peaks exhibit 
good agreement between the rescaled and non-scaled systems.
}
\label{fig1}
\end{figure}

Since the experimental discovery of graphene \cite{Novoselov04}, 
two-dimensional materials have been at the focus of intensive research in 
condensed-matter physics, among others because they bear great promise 
for technological applications, see, e.g., 
Refs.~\cite{Novoselov2012,Novoselov16}. With respect to spintronics 
applications \cite{Avsar20}, it could nevertheless be a disadvantage that 
bulk graphene is non-magnetic and one needs to resort to the enhanced 
density of states at the Fermi level close to defects or zigzag borders 
in order to drive magnetic instabilities (see Ref.~\cite{Yazyev2010} and 
references therein). Recently, a twist appeared in the field when 
superconducting and correlated insulating states were discovered in 
experiments on bilayer graphene where one layer is rotated with respect 
to the other by a so-called ``magic'' angle \cite{Fatemi18,Cao18}, see 
Fig.~\ref{fig1}(a) for an illustration of such a ``twisted'' honeycomb 
bilayer, Ref.~\cite{andrei2020graphene} for a summary of some recent 
developments, and Refs.~\cite{tbgI,tbgII,tbgIII,tbgIV,tbgV,tbgVI} for 
examples of resulting theoretical efforts. Even if the nature of the 
correlated insulating state in these systems remains under debate (see, 
e.g., 
Refs.~\cite{Senthil18,Kuroki18,Pizarro19,Roy19,MacDonald18,Zhang20,Cea20,Klebl20,Laksono19,Bultinck20,liu2019theories}), 
it is reminiscent of the textbook antiferromagnetic insulator that 
appears in the Hubbard model for strong on-site Coulomb interaction $U$ 
\cite{Fazekas99}. Indeed, the defining feature of the magic angles 
\cite{LopesdosSantos07,Trambly10,SuarezMorell10,Bistritzer11,Trambly12} 
is the emergence of flat minibands around the Fermi level such that the 
relative importance of intrinsic interactions in graphene is enhanced. It 
has been demonstrated experimentally that ferromagnetism emerges when a 
suitable number of electrons is doped into these flat bands 
\cite{Sharpe19}, a fact that might actually be a manifestation of the 
general phenomenon of flat-band ferromagnetism in the Hubbard model for 
suitable filling fractions \cite{pons2020flatband}.

%%%%%%%%%%%%%%%%%%%%%%%
%%%%%%%%%%%%%%%%%%%%%%%

Here we reexamine the one-band Hubbard model for twisted bilayer graphene 
(TBG) and demonstrate that magnetism occurs also in the charge-neutral 
(half-filled) system at low values of the on-site Coulomb interaction $U$, 
thus placing magnetic states, including an antiferromagnetic one, among 
the competitors for the instabilities in charge-neutral magic-angle 
twisted bilayer graphene.

We start from the tight-binding model of 
Refs.~\cite{Trambly10,Trambly12}. The resulting 
non-interacting band structure at the first magic angle 
$\theta=1.08^\circ$ is shown by the full blue line in Fig.~\ref{fig1}(b) 
and the corresponding total density of states (DOS) in Fig.~\ref{fig1}(c). 
The four flat minibands and the related strong enhancement of the DOS at 
$E_F$ are evident. On top of that, we add Coulomb interactions between the 
electrons in terms of a local on-site Hubbard interaction $U$. The 
resulting magnetic instabilities are then investigated by a combination of 
real-space static mean-field theory (MFT) \cite{Yazyev2010} and dynamical 
mean-field theory (DMFT) \cite{Georges1996,Marcin19,Thu2020}. As an 
alternative to MFT, one could determine the instabilities of the 
paramagnetic state with a random-phase approximation (RPA) analysis 
\cite{Klebl19}, and we present results from such an RPA analysis in 
appendix \ref{app:RPA}.

\section{Geometry of twisted bilayer graphene (TBG)}

Let us start by explaining the geometry of TBG in more detail. A single 
layer of graphene consists of carbon atoms arranged in a honeycomb lattice 
such that the unit cell includes two sites. We then construct a periodic 
commensurate bilayer structure parameterized by two integers $m$, $n$ 
using the method of 
Refs.~\cite{LopesdosSantos07,Trambly10,Trambly12,Santos12,Moon13}. $m$ and 
$n$ are coordinates with respect to the lattice vectors of a single 
graphene layer $\vect{a}_{1,2}=a(\sqrt{3},\pm1)/2$. The rotation angle for 
such a commensurate structure (moir\'e pattern) is then given by
\begin{equation}
\cos\theta=\frac{n^2+m^2+4mn}{2(n^2+m^2+mn)} \, ,
\label{eq:angle}
\end{equation}
and the fundamental vectors of the TBG superlattice are 
$\vect{t}_1=n\vect{a}_1+m\vect{a}_2$ and 
$\vect{t}_2=-m\vect{a}_1+(m+n)\vect{a}_2$. The number of atoms in the 
moir\'e cell is given by
\begin{equation}
N_c=4(n^2+m^2+mn) \, .
\label{eq:Nc}
\end{equation}
Figure~\ref{fig1}(a) shows the resulting moir\'e pattern for $(n,m)=(8,9)$ 
corresponding to a twist angle $\theta=3.89^\circ$ and $N_c = 868$ atoms 
in the moir\'e cell.

\section{Model Hamiltonian}
We start from the tight-binding model for the $p_z$ orbitals of the carbon 
atoms in charge-neutral TBG: $\hat{H}=\hat{H}_0+\hat{H}_{\rm int}$, where 
$\hat{H}_0$ is the single-electron Hamiltonian and $\hat{H}_{\rm int}$ is 
the electron-electron interaction. This leads to the one-band Hubbard 
model
\begin{equation}
 \hat{H}=\sum_{i,j,\sigma}t(\vect{r}_i;\vect{r}_j)\,\hat{d}_{i\sigma}^{\dagger}\hat{d}_{j\sigma}+U\sum_i\left(\hat{n}_{i\uparrow}-\frac12\right)\left(\hat{n}_{i\downarrow}-\frac12\right) \, ,
\label{eq1}
\end{equation}
where $\hat{d}_{i\sigma}^{\dagger}$ and $\hat{d}_{i\sigma}$ are the 
creation and annihilation operators of an electron with spin projection 
$\sigma=\{\uparrow,\downarrow\}$ at site $i$ and 
$\hat{n}_{i}=\sum_{\sigma}\hat{d}_{i\sigma}^{\dagger}\hat{d}_{i\sigma}$ is 
the total electron density at site $i$.  The hopping parameters 
$t(\vect{r}_i;\vect{r}_j)$ between two $p_z$ orbitals located at 
$\vect{r}_i$ and $\vect{r}_j$ are given in 
Refs.~\cite{Trambly10,Trambly12}. The second term in Eq.~(\ref{eq1}) 
describes the on-site Coulomb repulsion. The resulting non-interacting 
band structure ($U=0$) at the first magic angle $\theta=1.08^\circ$ is 
shown by the full blue line in Fig.~\ref{fig1}(b). This case corresponds 
to $(n,m)=(30,31)$ and thus to a moir\'e cell with $N_c = 11164$ sites. 
Dealing with such big unit cells will be challenging even for a one-band 
model and even within mean-field theory (MFT) and thus we will explore an 
idea of Ref.~\cite{Luis17} to reduce the numerical effort.

The precise non-interacting band structure depends not only on the 
geometry, but evidently also on the hopping parameters 
$t(\vect{r}_i;\vect{r}_j)$, and in particular the ratio between intra- and 
interlayer hopping. Let $\theta$ and $\theta'$ be the angles corresponding 
to two commensurate moir\'e structures and
\begin{equation}
\scaleFac=\frac{\sin{\frac{\theta'}{2}}}{\sin{\frac{\theta}{2}}} \, .
\label{eq:Lambda}
\end{equation}
Then the rescaling $t_0' = \scaleFac\,t_0$ of the nearest-neighbor 
intralayer hopping while keeping the interlayer hopping unchanged maps the 
low-energy band structure from the unprimed to the primed geometry 
\cite{Luis17}. The panels (b) and (c) of Fig.~\ref{fig1} illustrate this 
mapping for the first magic angle from $\theta=3.89^\circ$ to $\theta_{\rm 
eff}\equiv \theta' = 1.08^\circ$. Indeed, the dashed red line reproduces 
both the low-energy band structure and the density of states well at the 
expense of reducing the nearest-neighbor intralayer hopping from the 
physical value $t_0 = 2.7$\,eV~\cite{CastroNeto2009,Yazyev2010} to $t_0' 
\approx 0.75$\,eV, {\it i.e.}, modifying the high-energy physics. With 
different rescaling factors, {\it i.e.}, $t_0' \approx 0.90$\,eV and 
$1.02$\,eV, we can also model the angles $\theta = 1.30^\circ$ and 
$1.47^\circ$ in the $(n,m) = (25,26)$ and $(22,23)$ systems, respectively 
by the same effective $(n,m)=(8,9)$ system.

Ref.~\cite{Luis17} suggested that the on-site Coulomb interaction should 
scale in the same way as the intralayer hopping parameters, $U' = 
\scaleFac\,U$ although this is less evident than the rescaling of the 
hopping parameters, as we will also see in the results to be presented 
below.

In the following section \ref{secResc} we will first explore this 
rescaling trick in order to perform a detailed study using the case 
$(n,m)=(8,9)$ ($N_c = 868$). In section \ref{sec:NoScale} we will then 
check for some representative cases to what extent the conclusions do 
indeed apply to the unscaled system, including the first magic angle, {\it 
i.e.}, $(n,m)=(30,31)$ ($N_c = 11164$).

\section{Rescaled system}

\label{secResc}

In this section, we investigate the Hubbard model (\ref{eq1}) for twisted 
bilayer graphene (TBG) using rescaled interlayer hopping parameters, as 
outlined in the previous section. We will start with a systematic study 
using static MFT and then use a more sophisticated dynamical mean-field 
theory (DMFT) to argue that the findings of the simple MFT are
qualitatively correct even if there is a quantitative renormalization 
of the values of the on-site Coulomb interaction $U$.

\subsection{Static mean-field theory (MFT)}

\label{secRescMFT}

Static MFT is a well-established method to investigate the magnetism in graphene (see, e.g., chapter 3.1 of \cite{Yazyev2010}
and Refs.~\cite{Feldner2010,%FeldnerE,
Feldner2011,Luis17,Marcin19,Thu2020})
such that here we summarize only the essential features.
It amounts to the Hartree-Fock approximation of the interaction term in Eq.~(\ref{eq1}),
\begin{equation}
U\,n_{i\uparrow}n_{i\downarrow}\approx U\,\left(\langle n_{i\uparrow}\rangle \, n_{i\downarrow}
  + \langle n_{i\downarrow}\rangle \, n_{i\uparrow}
  - \langle n_{i\uparrow}\rangle \, \langle n_{i\downarrow}\rangle \right) \, ,
\label{eq:HF}
\end{equation}
where $\langle n_{i\sigma}\rangle$ is the average electron occupation 
number with spin $\sigma$ at site $i$. Note that the approximation 
(\ref{eq:HF}) decouples the operators for the two spin sectors and thus 
gives rises to a quadratic Hamiltonian in each of them where the other 
spin sector enters only via the site-dependent mean fields $\langle 
n_{i\sigma} \rangle$ that have to be determined self-consistently. We 
focus on charge-neutral TBG that has exactly one electron per site, {\it 
i.e.}, we work with the half-filled Hubbard model. A self-consistent 
solution is found iteratively, where in each step $N_c \times N_c$ 
matrices need to be diagonalized and an integral over the moir\'e 
Brillouin zone has to be calculated, that we approximate by a uniform grid 
of $\vect{k}$ points. We iterate this procedure until the maximum change 
of a density is below $10^{-6}$. Given the necessity to diagonalize a 
large number of moderately-sized matrices, even this elementary MFT 
approach becomes CPU-time intensive in the present situation. Some checks 
indicate that a $\vect{k}$-grid of at least $9\times 9$ points is required 
to eliminate artifacts of this discretization while more points do not 
change the conclusions. We therefore show results below that have been 
obtained for $9\times 9$ $\vect{k}$ points.

The RPA analysis that we present in appendix \ref{app:RPA} reveals 
different competing magnetic instabilities at different values of 
$\vect{q}$ for the present model. There is a periodic solution with an 
antiferromagnetic internal structure. The dominant instabilities are 
actually found at $\vect{q} \ne \vect{0}$, {\it i.e.}, they should have a 
larger unit cell than the twisted bilayer lattice, and they have a 
ferromagnetic structure inside a moir\'e cell. Motivated by the fact that 
the Hubbard model on a single honeycomb layer becomes antiferromagnetic at 
large $U$ \cite{Sorella1992}, we focus here on the antiferromagnetic 
mean-field solution. The RPA analysis of appendix \ref{app:RPA} predicts a 
critical value $U_c \approx 0.23\,t_0'$ for the antiferromagnetic state of 
the twisted bilayer system with $\theta_{\rm eff}=1.08^\circ$, an order of 
magnitude below the critical value of a single layer, that for 
nearest-neighbor hopping is known to be $U^{\rm MFT}_c/t\approx2.23$ 
\cite{Sorella1992}.

%%%%%%%%%%%%%%%%%%%%%%%
%%%%%%%%%%%%%%%%%%%%%%%
%%%%%%%%%%%%%%%%%%%%%%%
\begin{figure}[t!]
\centering
\includegraphics[width=0.78\linewidth]{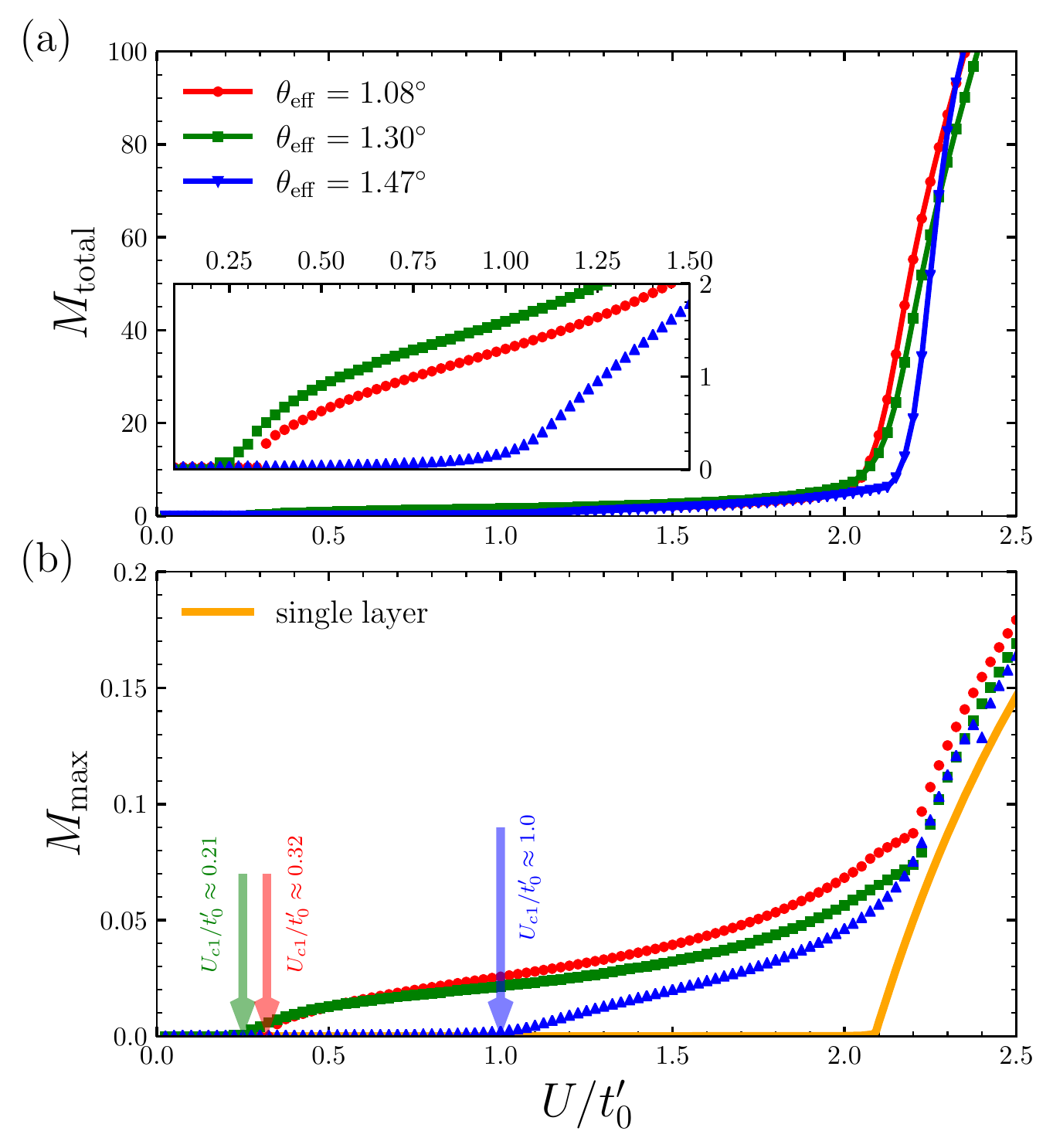}
\caption{MFT results for the magnetization of the rescaled twisted bilayer 
system as a function of on-site Coulomb interaction $U/t_0'$. Panels (a) 
and (b) show the total magnetization per effective $N_c=868$ moir\'e cell 
and its maximum, respectively. For comparison, results for a single 
graphene layer with the same intralayer hopping parameters are also shown 
in panel (b). In panel (b), red, green, and blue arrows mark the critical 
points for $\theta_{\rm eff}=1.08^\circ$, $\theta_{\rm eff}=1.30^\circ$, 
and $\theta_{\rm eff}=1.47^\circ$, respectively.
}
\label{fig2}
\end{figure}
%%%%%%%%%%%%%%%%%%%%%%%
%%%%%%%%%%%%%%%%%%%%%%%
%%%%%%%%%%%%%%%%%%%%%%%

Figure~\ref{fig2} shows MFT results for the total magnetization per 
moir\'e cell and its maximum value, defined as
\begin{eqnarray}
M_{\rm total} &=&\sum_i^{N_c} \abs{m_z(\vec{r}_i)} \, , \qquad
m_z(\vec{r}_i) = \frac{\langle n_{i\uparrow}\rangle-\langle n_{i\downarrow}\rangle}2 \,,
\label{eq:Mtot}\\
M_{\rm max} &=& {\rm max}\{\abs{m_z(\vec{r}_1)},\cdots,\abs{m_z(\vec{r}_{N_c})}\} \, ,
\label{eq:Mmax}
\end{eqnarray}
respectively.
We first focus on the first magic angle $\theta_{\rm eff}=1.08^\circ$ (red 
data in Fig.~\ref{fig2}). Here, we find a small albeit finite 
magnetization for values of $U/t$ as low as
\begin{equation}
U_{c1,{\rm MFT}}^{1.08^\circ}/t_0' \approx 0.32 \, .
\label{eq:Uc1MFT}
\end{equation}
We note that convergence is delicate close to $U_{c1,{\rm MFT}}$ and 
sensitive to the chosen $\vect{k}$ grid. The result (\ref{eq:Uc1MFT}) 
should thus be considered as an upper bound. Thus, we conclude that this 
value is consistent with the prediction of the RPA analysis of appendix 
\ref{app:RPA}. Given that the magnetization for these small values is due 
to the four flat minibands and that there is a low number of associated 
states (4 per moir\'e cell), the total magnetization (\ref{eq:Mtot}) for 
small values of $U$ is small and thus seen more clearly in the inset of 
Fig.~\ref{fig2}(a) than in the main panel. Indeed, for $U/t_0' \lesssim 
1.5$, the total magnetization per moir\'e cell remains below $2 = 4 \cdot 
1/2$, consistent with it coming mainly from the four flat minibands.

An alternative perspective is given by the maximum magnetization 
(\ref{eq:Mmax}) that is shown in Fig.~\ref{fig2}(b). Here, one can firstly 
observe the onset of magnetization around $U_{c1}$ more clearly than in 
the main panel of Fig.~\ref{fig2}(a). For comparison, the main panel of 
Fig.~\ref{fig2}(b) also includes the result for a single layer with the 
same intralayer hoppings as in the twisted bilayer system. One observes 
firstly that additional long-range hoppings within each layer reduce the 
critical value slightly to $U^{\rm MFT}_c /t\approx2.09$ as compared to 
the nearest-neighbor result $U^{\rm MFT}_c /t\approx2.23$ 
\cite{Sorella1992}. In the region $U/t_0' \gtrsim 2$, the magnetization of 
the bilayer system is slightly enhanced with respect to the single-layer 
case, as might be expected thanks to the additional intralayer couplings. 
However, the transition to full magnetization necessarily involves AB and 
BA stacking regions (see Fig.~\ref{fig1}(a)) that are geometrically 
frustrated. Consequently, one expects a complex magnetic state in this 
transition region. A full analysis of the transition to a fully magnetized 
system is beyond the scope of the present work, but we note that 
convergence is delicate also in this second transition region, as 
exemplified by the outlier at $U/t_0' = 2.4$ in the $\theta_{\rm 
eff}=1.47^\circ$ data.

The most important finding in the present context is that magnetism arises 
in the effective twisted bilayer model at the magic angle for Coulomb 
interactions $U$ that are an order of magnitude smaller than for decoupled 
single graphene layers. It should be noted that the $\vect{q}=\vect{0}$ 
magnetic solution considered here only corresponds to a local, but not the 
global minimum of the energy such that the true critical value of 
$U_{c1,{\rm MFT}}^{1.08^\circ}$ is probably even smaller than the result 
(\ref{eq:Uc1MFT}) ($U_{c1,{\rm MFT}}^{1.08^\circ} \approx 0.15\,t_0'$ 
according to the RPA analysis of appendix \ref{app:RPA}).

Figure~\ref{fig2} also includes two examples for larger twist angles 
$\theta_{\rm eff}=1.30^\circ$ and $1.47^\circ$ (green and blue data, 
respectively). Many of the preceding remarks also apply to these two cases 
such that we focus on their peculiarities. Remarkably, the case 
$\theta_{\rm eff}=1.30^\circ$ yields an even smaller $U_{c1,{\rm 
MFT}}^{1.30^\circ}/t_0' \approx 0.21$ than for $\theta_{\rm eff} = 
1.08^\circ$. Actually, while the velocity at the K point only vanishes at 
the first magic angle $\theta = 1.08^\circ$, the minibands have a very 
small bandwidth over the entire range until $\theta=1.30^\circ$. However, 
when one goes to $\theta_{\rm eff}=1.47^\circ$, the critical value of the 
onsite Coulomb repulsion increases to $U_{c1,{\rm MFT}}^{1.48^\circ}/t_0' 
\approx 1.0$. This is still significantly smaller than the critical value 
of a single layer $U^{\rm MFT}_c /t\approx2.09$, but clearly larger than in 
the two other cases, as expected for minibands close to the Fermi level 
that now have both a finite Fermi velocity and a significant bandwidth.

%%%%%%%%%%%%%%%%%%%%%%%
%%%%%%%%%%%%%%%%%%%%%%%
%%%%%%%%%%%%%%%%%%%%%%%
\begin{figure}[t!]
\centering
\includegraphics[width=0.9\linewidth]{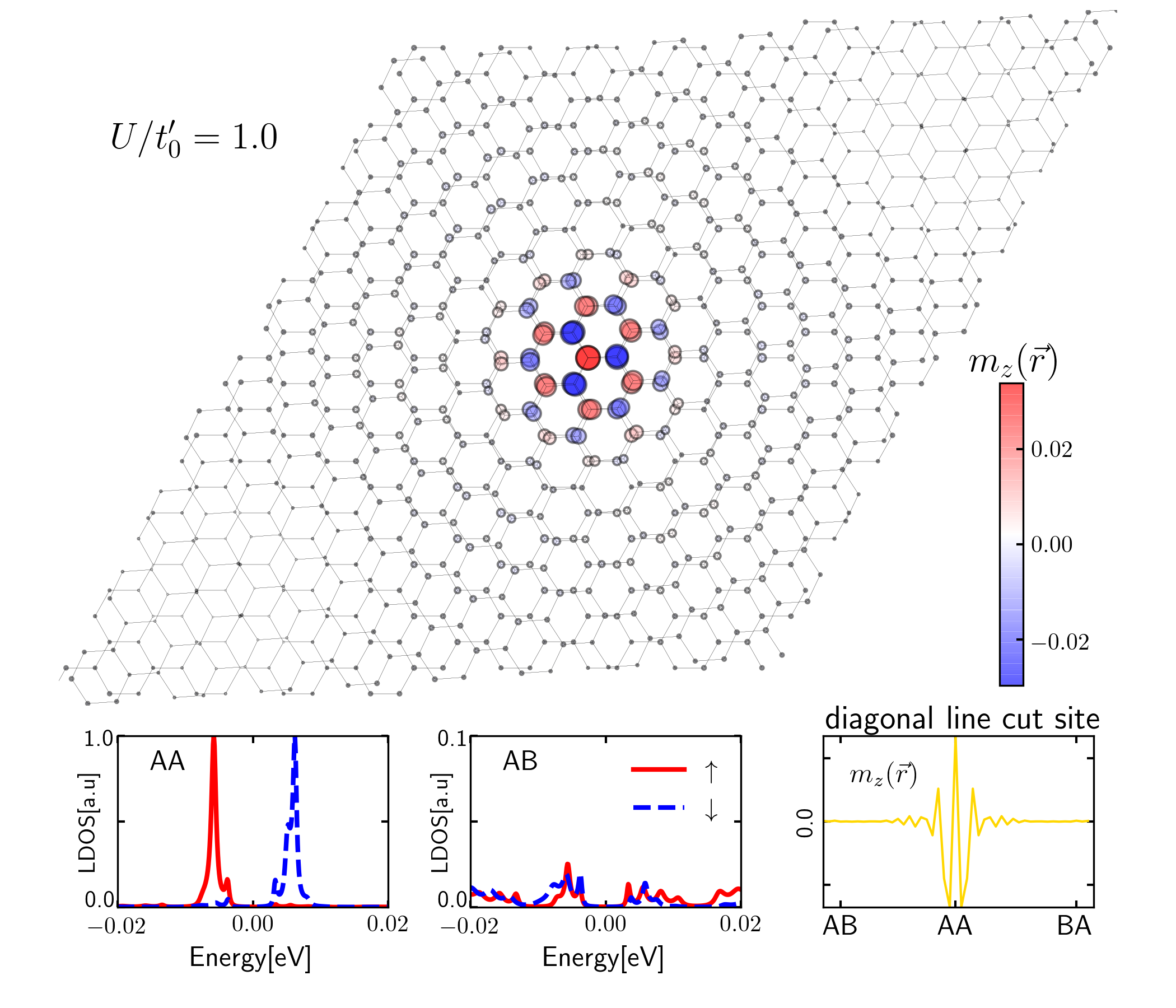}
\caption{Top panel: MFT result for the spatial magnetization profile of a 
rescaled twisted bilayer with $\theta_{\rm eff}=1.08^\circ$, and the 
on-site Coulomb interaction $U/t_0'=1$. The bottom panels show the local 
density of states (LDOS) in the AA and AB regions for both spin 
projections (left two panels), and a diagonal line cut of the local 
magnetic moment (right panel).}
\label{fig3}
\end{figure}
%%%%%%%%%%%%%%%%%%%%%%%
%%%%%%%%%%%%%%%%%%%%%%%
%%%%%%%%%%%%%%%%%%%%%%%

For a more detailed discussion of the magnetic state found above $U_{c1}$ 
but before the system becomes completely magnetic, we show in 
Fig.~\ref{fig3} results for the $\theta_{\rm eff}=1.08^\circ$ system and a 
representative value of the on-site Coulomb interaction $U/t_0'=1$. The 
top panel shows the spatial structure of the magnetization pattern that we 
find to be localized in the AA stacking region. Thus, in this region the 
magnetic state of the twisted bilayer system resembles that of AA stacked 
bilayer graphene, but at a significantly lower value of $U$ than would be 
required for the simple AA system to become magnetic. A different 
perspective of this magnetic pattern is provided by the lower right panel 
of Fig.~\ref{fig3} that presents a diagonal line cut of the magnetization. 
The lower left two panels of Fig.~\ref{fig3} show the spin-resolved local 
density of states (LDOS) in the AA and AB stacking regions. Interestingly, 
in the AA region one finds two peaks in the LDOS at low energies that are 
absent in the AB stacking region. The presence of these peaks correlates 
with the magnetic state, thus rendering scanning tunneling microscopy 
(STM) experiments a promising candidate for the detection of such a 
magnetic state.

\subsection{Dynamical mean-field theory (DMFT)}

Even though MFT has been shown to be remarkably successful to 
qualitatively describe static \cite{Feldner2010} %,FeldnerE}
and dynamic properties \cite{Feldner2011} in the semi-metallic phase
of single-layer graphene, it is known to become quantitatively less
accurate for larger values of $U$. For example, the transition to the
antiferromagnetic insulator in the nearest-neighbor hopping case is
found at $U^{\rm MFT}_c /t\approx\,2.23$ in MFT \cite{Sorella1992} while
more sophisticated and accurate methods place it at a larger
$U_c/t\approx3.8$ \cite{Sorella2012,Hassan2013,Assaad2013,Hirschmeier2018}.

DMFT \cite{Georges1996} takes local charge fluctuations into account and 
thus improves the quantitative treatment of the on-site Hubbard 
interaction. Indeed, already single-site DMFT shifts the estimate of the 
critical point to the range $U_c^{\rm DMFT}/t = 3.5, \ldots, 3.7$ 
\cite{Marcin19}, {\it i.e.}, remarkably close to the most accurate 
estimates \cite{Sorella2012,Hassan2013,Assaad2013,Hirschmeier2018}. 
Following previous work, we employ here a real-space version of DMFT 
\cite{Thu2020}. DMFT maps the lattice Hamiltonian Eq.~(\ref{eq1}) onto a 
set of quantum impurity problems via the local Green's function for site 
$i$ inside the moir\'e supercell~\cite{Georges1996}
\begin{equation}
G_{i\sigma}(z)=\int {\rm d}k
\, \left(z\mathbb{I}-\hat{H}_0(\vect{k})-\mathbf{\Sigma}^r_\sigma(z)\right)^{-1}_{i,i}\,.
\label{eq2}
\end{equation}
Here $\hat{H}_0$ is the single-particle part of Eq.~(\ref{eq1}). The main approximation is that
the local self-energy matrix for spin projection $\sigma$, $\mathbf{\Sigma}^r_{\sigma}(z)$, that
plays the role of a dynamical mean field, depends only on frequency $z$, but not on
momentum $\vect{k}$. Eq.~(\ref{eq2}) can be used to define a collection of $N_c$ single-impurity Anderson models, that we solve here
with the numerical renormalization group (NRG)~\cite{Wilson1975,Krishna1980,Bulla2008} and iterate until self-consistency is reached~\cite{Robert2014,Robert2015}. We refer to Refs.~\cite{Marcin19,Thu2020} for details on the procedure and just mention two peculiarities
for the present case. Firstly, Eq.~(\ref{eq2}) requires evidently a combination of integration over the moir\'e Brillouin zone while
at the same time solving coupled problems for the $N_c$ atoms inside the moir\'e supercell. Secondly, even if the band structure of
Fig.~\ref{fig1}(b), (c) is almost particle-hole symmetric, there is no strict particle-hole symmetry in the present case in contrast to previous work~\cite{Marcin19,Thu2020}. Consequently, the chemical potential needs to be adjusted appropriately during each iteration in order to ensure half filling.
Since the chemical potential enters into Eq.~(\ref{eq2}) in a non-linear fashion, this renders the numerical problem even more challenging,
thus limiting DMFT not only to the rescaled system, but also the number of $U$-values considered.

%%%%%%%%%%%%%%%%%%%%%%%
%%%%%%%%%%%%%%%%%%%%%%%
%%%%%%%%%%%%%%%%%%%%%%%
\begin{figure}[t!]
\centering
\includegraphics[width=0.78\linewidth]{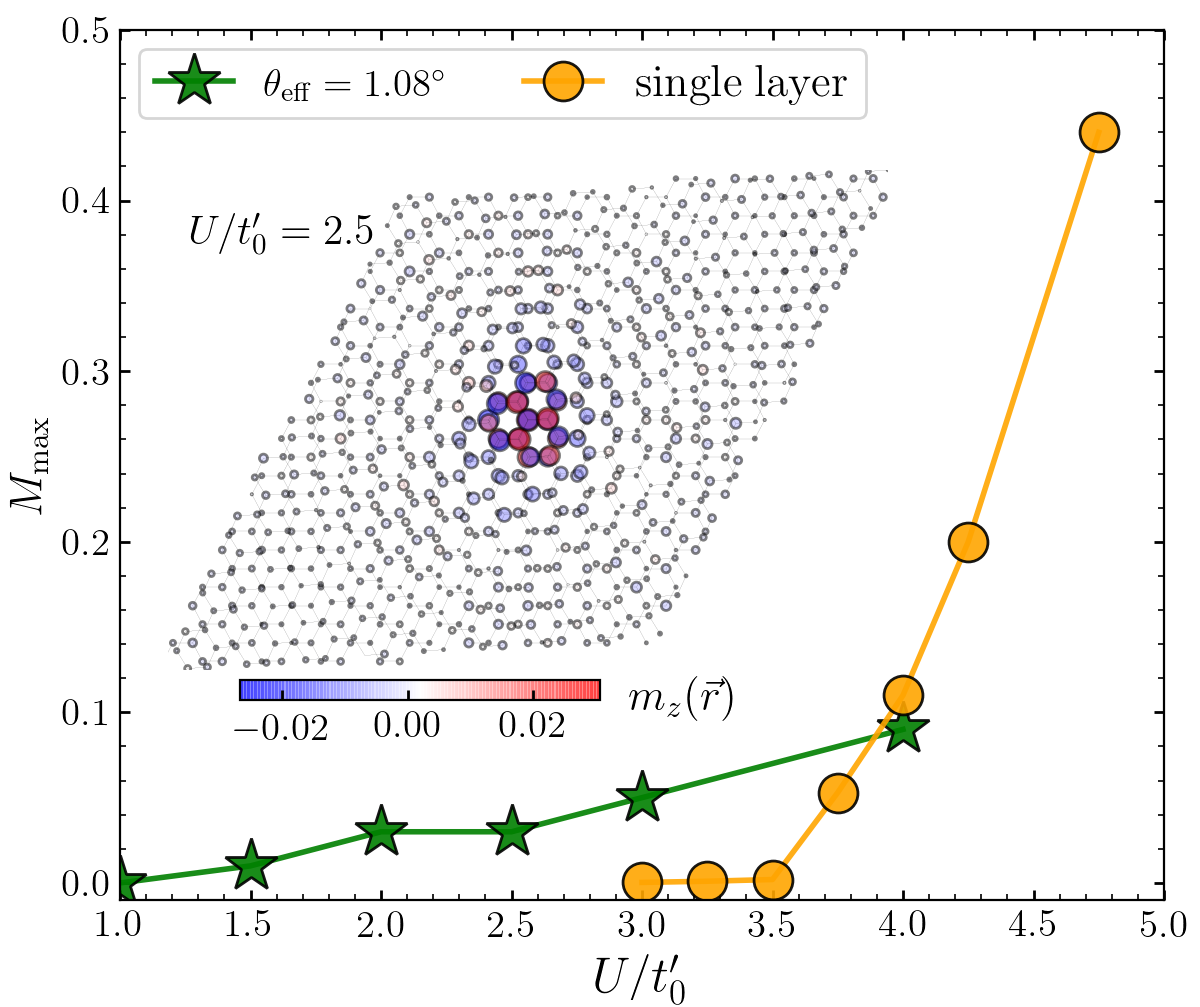}
\caption{DMFT results for the magnetization versus Hubbard interaction $U/t_0'$ for the rescaled system with
$\theta_{\rm eff}=1.08^\circ$. For comparison, results for a single graphene layer are also shown.
Lines are guides to the eye.
The inset shows the spatial magnetization for $U/t_0'=2.5$.}
\label{fig_DMFT}
\end{figure}
%%%%%%%%%%%%%%%%%%%%%%%
%%%%%%%%%%%%%%%%%%%%%%%
%%%%%%%%%%%%%%%%%%%%%%%

Figure \ref{fig_DMFT} presents some DMFT results for the rescaled system 
with $\theta_{\rm eff}=1.08^\circ$, {\it i.e.}, $N_c = 868$. Comparison of 
the DMFT results for the magnetization versus $U/t_0'$ in 
Fig.~\ref{fig_DMFT} with the MFT results of Fig.~\ref{fig2} shows 
qualitatively similar behavior. At a technical level, the DMFT results are 
a bit more noisy. This is due to the logarithmic frequency discretization 
inherent to NRG~\cite{Bulla2008}, to DMFT being generally numerically more 
expensive, and in particular the difficulty to adjust the chemical 
potential appropriately. Nevertheless, the main quantitative difference 
remains that the critical $U_c$ of a single layer is pushed to larger 
values as compared to simple MFT, and so is the phenomenon of a 
magnetization arising in the AA stacking region of the twisted bilayer 
system. Nevertheless, also DMFT clearly detects a magnetization in the 
twisted system for values of the local Coulomb interaction down to 
$U/t_0'=1$, amounting to a reduction of the critical value as compared to 
the single-layer system by at least a factor $3.5$ at $\theta_{\rm 
eff}=1.08^\circ$. The inset of Fig.~\ref{fig_DMFT} shows an example of the 
spatial magnetization pattern. This is again very similar to the MFT 
result shown in the main panel of Fig.~\ref{fig3}, just the value of $U/t$ 
is renormalized to larger values, namely from $1$ for the MFT example to 
$2.5$ of the DMFT example. Note that $U/t_0'=2.5$ would give rise to a 
bulk magnetic state within MFT while the DMFT result in the inset of 
Fig.~\ref{fig_DMFT} is still clearly localized in the AA stacking region. 
Overall, DMFT confirms the qualitative conclusions derived from MFT; it 
just provides a quantitatively more accurate account of the local Coulomb 
interaction $U$.

\section{Non-scaled system}

\label{sec:NoScale}

%%%%%%%%%%%%%%%%%%%%%%%
%%%%%%%%%%%%%%%%%%%%%%%
%%%%%%%%%%%%%%%%%%%%%%%
\begin{figure}[t!]
\centering
\includegraphics[width=0.78\linewidth]{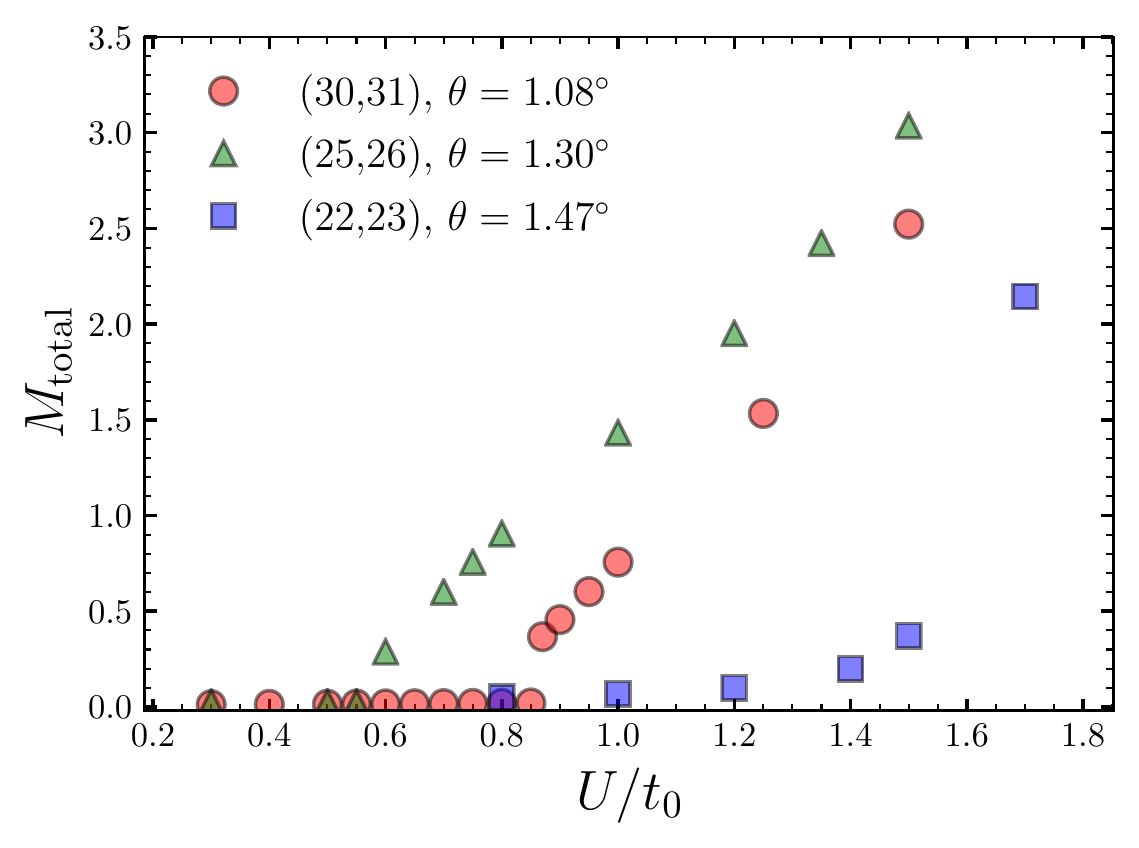}
\caption{MFT results for the magnetization of the non-scaled twisted
bilayer system as a function of $U/t_0$ at rotation angles
$\theta=1.08^\circ$, $\theta=1.30^\circ$, and $\theta=1.47^\circ$,
respectively.
}
\label{fig2n}
\end{figure}
%%%%%%%%%%%%%%%%%%%%%%%
%%%%%%%%%%%%%%%%%%%%%%%
%%%%%%%%%%%%%%%%%%%%%%%

We will now present some results for the non-scaled system. The scaling 
trick has allowed us to apply the quantitatively more accurate DMFT, but 
the size of the moir\'e cells of the non-scaled systems will exceed those 
accessible to DMFT such that we focus on static MFT in the present 
section. We use the same parameters as in section \ref{secRescMFT} 
(convergence criterion $10^{-6}$, $9\times 9$ $\vect{k}$-grid).

Figure~\ref{fig2n} shows MFT results for the total magnetization per 
moir\'e cell as a function of $U/t_0$ at rotation angles 
$\theta=1.08^\circ$, $\theta=1.30^\circ$, and $\theta=1.47^\circ$. The 
corresponding moir\'e cells contain $N=11164$, $7804$, and $6076$ carbon 
atoms, respectively. At first sight, the behavior is very similar to that 
found in the inset of Fig.~\ref{fig2}(a) for the rescaled system (the 
smaller number of data points is due to the significantly enhanced 
computational effort). In particular, $M_{\rm total} \lesssim 2$ remains 
true for most values of $U/t_0$ shown in Fig.~\ref{fig2n}, in agreement 
with again the magnetism beging due to the four flat minibands that are 
closest to the Fermi level.

The key items are the values of the critical Coulomb interaction that
one may estimate as
$U_{c1,{\rm MFT}}^{1.08^\circ}/t_0 \approx 0.85$,
$U_{c1,{\rm MFT}}^{1.30^\circ}/t_0 \approx 0.55$, and
$U_{c1,{\rm MFT}}^{1.47^\circ}/t_0 \approx 1$
with a particularly large uncertainty on the last result given the
very slow onset of magnetization for $\theta=1.47^\circ$.
According to Ref.~\cite{Luis17}, in the given normalization,
these values should correspond to those found in the rescaled system.
This works out more or less for the case $\theta=1.47^\circ$ where
in both cases, the critical $U/t$ ratio is close to $1$. However,
the values for $U_{c1,{\rm MFT}}^{1.08^\circ}$ and $U_{c1,{\rm MFT}}^{1.30^\circ}$
in the non-scaled system are bigger than those we might have expected from the
rescaled case. Indeed, the order of the discrepancy corresponds to another factor
$\scaleFac$ such that $U$ scales with $\scaleFac^2$ and not just with $\scaleFac$.
A possible interpretation of this observation is the following:
$\scaleFac$ actually also appears in the scaling of the linear
length~\cite{Luis17}. Now the magnetic instability at the angles
$\theta=1.08^\circ$ and $1.30^\circ$ is related to a state localized
in the AA region, see, e.g., top panel of Fig.~\ref{fig3}. Thus, the
area of the relevant spatial region scales with $\scaleFac^2$, accordingly
the number of contributing local on-site repulsions also scales
with $\scaleFac^2$ such that $U$ should also scale with $\scaleFac^2$
rather than $\scaleFac$ in the cases where the physics is controlled by
localized states.

In spite of this additional factor, it remains true that $U_{c1,{\rm 
MFT}}^{1.30^\circ}/t_0 \approx 0.55 < U_{c1,{\rm MFT}}^{1.08^\circ}/t_0 
\approx 0.85$, and that there is still a significant reduction by factors 
of 4 respectively 3 with respect to the critical value $U_c$ for a single 
graphene layer. In light of the preceding observations, we suggest that 
not only the non-interacting bandwidth, but also the size of the moir\'e 
cell matter. While both the $\theta=1.08^\circ$ and $1.30^\circ$ bilayers 
have a small bandwidth, the moir\'e cell of the latter is smaller, and 
this appears to result in a smaller critical value of $U_{c1}$. The size 
of the moir\'e cell is smallest for $\theta=1.47^\circ$ among the three 
cases studied, but the value of $U_{c1}$ is biggest, most likely because 
in this case the minibands closest to the Fermi energy are no longer flat. 
Nevertheless, even in this case one observes emergence of magnetism for 
values of $U$ that are about a factor 2 smaller than would be needed to 
render a single layer antiferromagnetic.

%%%%%%%%%%%%%%%%%%%%%%%
%%%%%%%%%%%%%%%%%%%%%%%
%%%%%%%%%%%%%%%%%%%%%%%
\begin{figure}[t!]
\centering
\includegraphics[width=\linewidth]{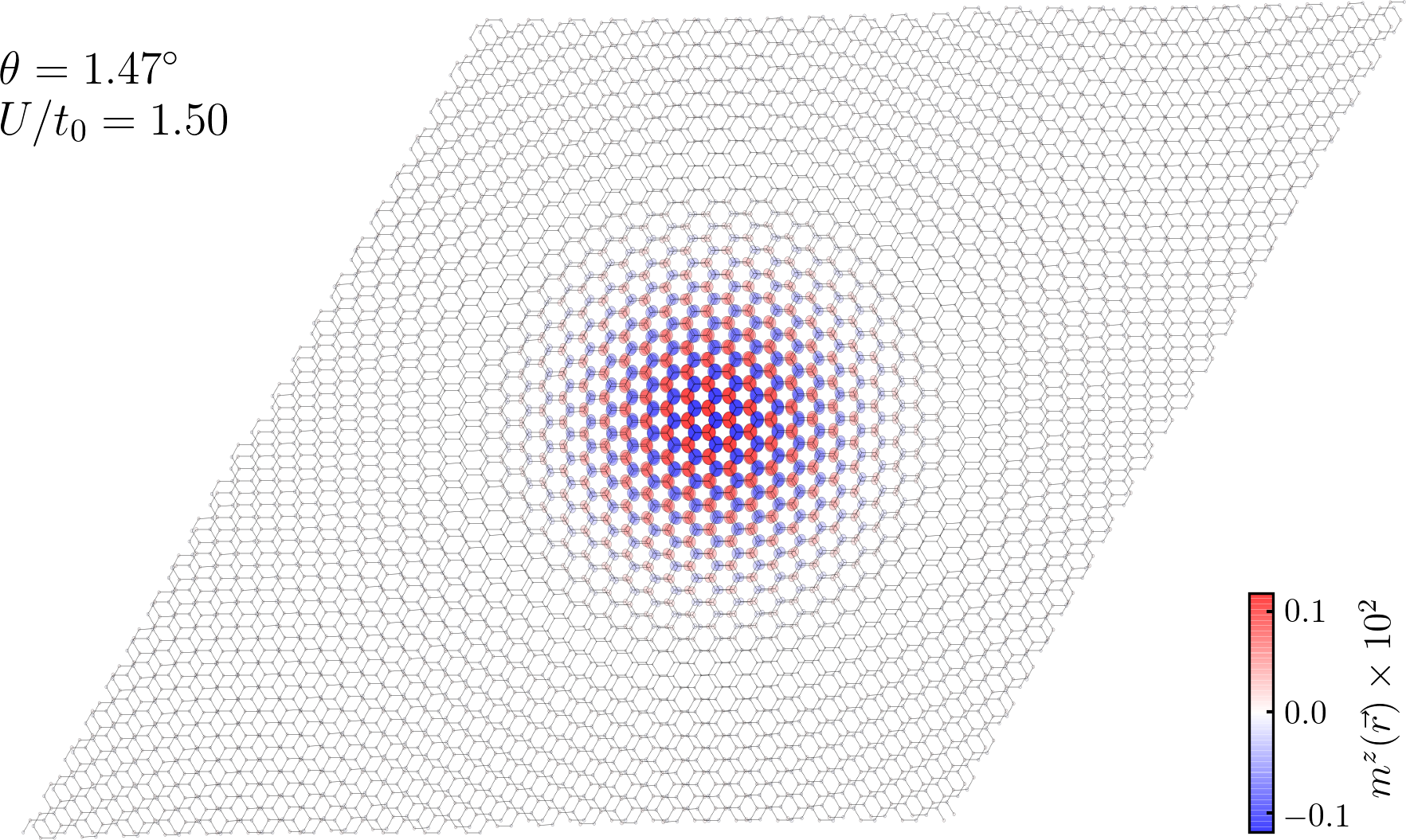}\\[3mm]
\includegraphics[width=\linewidth]{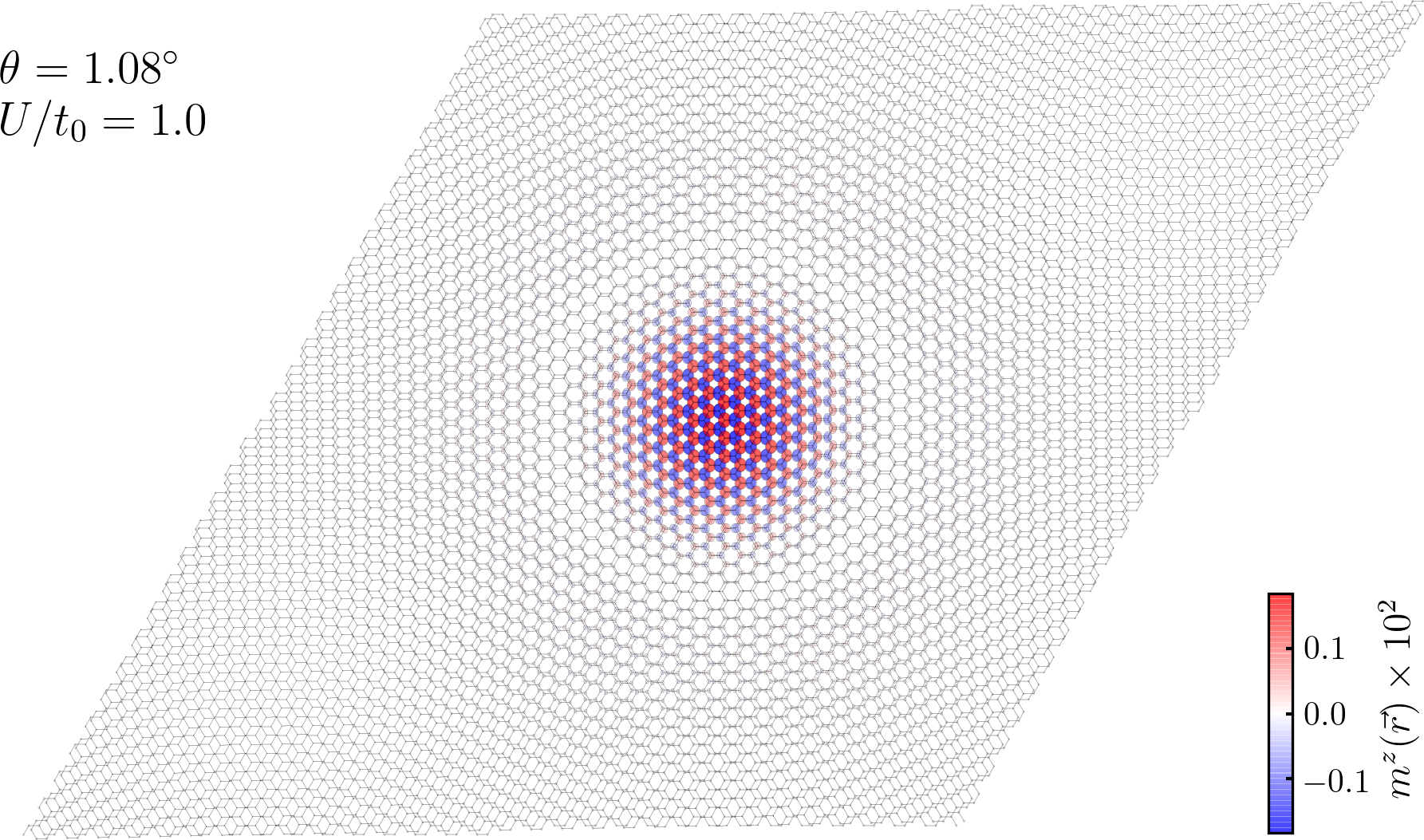}
\caption{The top and bottom panels show the spatial magnetization profile of non-scaled systems at $\theta=1.47^\circ$
and $\theta=1.08^\circ$, respectively.
The corresponding numbers of atoms in the unit cell are $N_c=6076$ and $N_c=11164$.}
\label{fig5}
\end{figure}
%%%%%%%%%%%%%%%%%%%%%%%
%%%%%%%%%%%%%%%%%%%%%%%
%%%%%%%%%%%%%%%%%%%%%%%

To conclude this discussion, let us have a closer look at the spatial 
structure of the resulting magnetic states. Figure \ref{fig5} shows the 
spatial magnetization profile for non-scaled moir\'e unit cells with 
angles $\theta=1.47^\circ$ and at the first magic angle 
$\theta=1.08^\circ$. For illustration purposes, we consider a value of $U$ 
just above the first critical point $U_{c1}$, {\it i.e.} $U/t_0=1.50$ and 
$1.00$, respectively. Like for the recaled system shown in the top panel 
of Fig.~\ref{fig3}, we find an antiferromagnetic pattern that is localized 
in the AA region. However, thanks to the improved spatial resolution, we 
can now observe a clearer separation of the magnetic regions between 
neighboring cells. For slightly larger $U$, the magnetic region grows, but 
the structure remains qualitatively similar as in Fig.~\ref{fig5}.

\section{Conclusions and perspectives}

We have investigated the onset of magnetism in charge-neutral 
``magic-angle'' twisted bilayer graphene with numerical real-space static 
and dynamical mean-field approaches. In the rescaled system we found that 
localized magnetic states appear in the twisted bilayer system for values 
of the on-site Coulomb repulsion $U$ that are an order of magnitude 
smaller than those needed to render a single layer magnetic. We then 
showed that the non-scaled system exhibits qualitatively similar behavior. 
The reduction is less impressive (up to a factor 4 in the cases 
investigated), but still remarkable. We note that this finding is 
consistent with a recent diagrammatic real-space mean-field study 
\cite{gonzalez2021magnetic} that focused on two selected values of $U$.

The rescaling proposed in Ref.~\cite{Luis17} actually reproduces the flat 
minibands close to the Fermi level very well, compare Fig.~\ref{fig1}. Our 
results therefore demonstrate that the band structure is not the only 
factor that matters. Indeed, the corresponding states are localized in AA 
stacking regions. This suggests a scaling of the critical $U_c$ with area 
rather than linear size, as is indeed roughly consistent with our findings 
for $\theta = 1.08^\circ$ and $1.30^\circ$. A more quantitative analysis 
would involve computation of the Coulomb matrix elements with respect to 
the Wannier functions 
\cite{Koshino18,Kang18,Senthil18,tbgIII,Vafek19,davydov2020} of the 
rescaled and non-scaled systems, respectively. However, such an analysis 
goes beyond the scope of the present work.

A side effect of the observation that the spatial extent of the localized 
states also matters is that smaller unit cells favor magnetism over bigger 
ones. Indeed, we find onset of magnetism for $\theta = 1.30^\circ$ for 
smaller values of $U_c$ than for the first magic angle $\theta = 
1.08^\circ$. The system with $\theta = 1.47^\circ$ has an even smaller 
unit cell than that with $\theta = 1.30^\circ$, but at this larger angle 
there is no longer any really flat band close to the Fermi level such that 
here the value of $U_c$ is found to be larger. A related point is that 
magic angles are usually defined via a vanishing Fermi velocity 
\cite{LopesdosSantos07,Trambly10,SuarezMorell10,Bistritzer11,Trambly12} 
while in fact it may be more relevant that the entire minibands are 
narrow. Indeed, the latter criterion is satisfied over the entire range 
$\theta = 1.08\ldots 1.30^\circ$ such that the smaller unit cell can then 
give rise to a lower $U_c$ at the upper boundary of this range of angles.

It should be noted that in our mean-field investigations we have focussed 
on antiferromagnetic solutions that are periodic over moir\'e cells. 
However, the RPA analysis of appendix \ref{app:RPA} suggests that there 
are other competing instabilities, and indeed the mean-field 
self-consistency loop sometimes converges to other solutions. In 
particular, the true lowest-energy state might be modulated in real space 
and exhibit an internal ferromagnetic structure, like in the case of an 
electric bias between the two layers \cite{Luis17}. Should this indeed be 
the case, this can only further reduce the value of the $U_c$ for the 
onset of magnetism such that our estimates are in fact upper bounds. The 
main conclusion that twisting leads to a significant reduction of the 
critical $U_c$ for the appearance of magnetism is thus unaffected by the 
assumptions on the nature of the ground state.

Another point to note is that we find a stronger reduction of the critical 
interaction $U_c$ at charge neutrality than a previous RPA investigation 
\cite{Klebl19}. This difference can be traced to a different tight-binding 
model at the starting point. Indeed, the authors of Ref.~\cite{Klebl19} 
have implemented the corrugation of Ref.~\cite{Koshino18} that takes a 
modulation of the distance between the two layers in different stacking 
regions into account. However, other factors may also be relevant in an 
experiment such as strain when the bilayer is deposited on a substrate. In 
the same spirit, Coulomb interactions should actually be long-range 
\cite{klebl2020}, at least for free-standing bilayers, since atomically 
thin layers cannot screen the Coulomb repulsion between electrons. Still, 
screening will depend on the actual substrate and may thus depend on the 
exact experimental conditions. Even other factors such as spin-orbit 
interactions that are sufficiently weak to be usually negligible in 
graphene may matter in the present situation given the significant 
reduction of the kinetic energy scale in the twisted bilayer system. Thus, 
which of several competing instabilities finally wins in an experimental 
realization may depend on a number of factors; here we have simply 
demonstrated that a magnetic instability (possibly an antiferromagnetic 
one) is one of the competitors in charge-neutral twisted bilayer graphene.

The macroscopic magnetization of a ferromagnetic state can be detected, 
e.g., via the Hall effect \cite{Sharpe19}. Antiferromagnetic or almost 
ferromagnetic, but modulated spiral states are more difficult to detect 
experimentally since they do not give rise to a macroscopic moment. In 
bulk systems, one would resort to (neutron) scattering to detect such 
states, but in the present nanoscopic setting this may not be feasible. 
The best option may thus be scanning tunneling spectroscopy (STS) 
experiments \cite{Xie19} in order to detect the corresponding 
characteristic features in the local density of states (see lower panels 
of Fig.~\ref{fig3}). In fact, the corresponding signatures might already 
have been observed in recent STS experiments 
\cite{Kerelsky2019,Choi2019,Jiang2019}. However, the latter samples are 
subject to heterostrain \cite{Huder2018,Mesple2020} which also gives rise 
to a splitting in the electronic density of states. An unambiguous 
detection of a magnetic state would thus require a detailed investigation 
of the variation of the tunneling spectrum with the different stacking 
regions.

Returning to theoretical questions, an alternative approach would be via 
low-energy continuum models in the spirit of Ref.~\cite{Bistritzer11}. One 
reason why we have rather used the rescaled model \cite{Luis17} here is 
that, as illustrates Fig.~\ref{fig1}, it reproduces the band structure 
well over a wide range of energies and not just the flat minibands close 
to the Fermi level. However, in the range of intermediate values of $U/t$ 
where mainly the flat minibands contribute to the magnetism, effective 
low-energy models would have the advantage of being more amenable to 
numerical approaches 
\cite{tbgVI,Liao21,Wilhelm21,pahlevanzadeh2021,potasz2021} such that we 
suggest the investigation of magnetism by such methods as a topic for 
further studies.

A further interesting issue that goes beyond questions accessible to 
low-energy effective models would be the full phase diagram of the twisted 
bilayer systems up to larger values of $U/t$. Indeed, the results 
underlying Fig.~\ref{fig2} suggest that there is no single simple 
transition to a bulk magnetized system, but that this transition actually 
proceeds via several intermediate states in the region $U/t_0' \approx 2$. 
Given that magnetic interactions in the AB and BA stacking regions are 
geometrically frustrated (compare Fig.~\ref{fig1}(a)), even the magnetic 
state in the Heisenberg limit $U/t \to \infty$ is far from obvious.

\section*{Acknowledgements}
This work was supported by the ANR project J2D ``Atomically sharp 
junctions based on stacked 2D materials: new building blocks for the 
electronics'' and the Paris//Seine excellence initiative. R.P.\ is 
supported by JSPS, KAKENHI Grant No.\ JP18K03511. MFT calculations have 
been performed at the Centre De Calcul (CDC), CY Cergy Paris Universit\'e
and using HPC resources from GENCI-IDRIS (Grant No.\ 910784). We thank
Y.\ Costes and B.\ Mary, CDC, for computing assistance. DMFT computations
in this work have been done using the facilities of the Supercomputer Center 
at the Institute for Solid State Physics, University of Tokyo.

\begin{appendix}

\section{Noninteracting susceptibility of the rescaled system}

\label{app:RPA}

In this appendix, we provide an RPA analysis of the noninteracting 
susceptibility that is similar in spirit to Ref.~\cite{Klebl19}. However, 
here we focus on the rescaled system with $\theta_{\rm eff}=1.08^\circ$.

We adopt the multiorbital RPA approach to study the instability of the 
paramagnetic state~\cite{Maier09,Liu18}.

The multiorbital spin susceptibilities tensor can be expressed in terms of 
the Matsubara spin-spin correlation function:
\begin{equation}
\big[\chi(\vect{q},\omega)\big]_{st}
 =\frac{1}{3}\int\limits_0^\beta~{\rm d}\tau \, {\rm e}^{i\omega\tau}  \Big< T_\tau \hat{\vect{S}}_s(\vect{q},\tau)\hat{\vect{S}}_t(-\vect{q},0)\Big>
\end{equation} 
with the Matsubara frequency $\omega$, the imaginary time $\tau$ and  spin operators $\hat{\vect{S}}$ at orbitals $s$, $t$. 
The noninteracting (zero-order) susceptibility is just a simple bubble diagram involving two Green's functions.
Using the spectral representation of the Green's functions, this can be expressed as
\begin{equation}
\big[\chi_0(\vect{q},i\omega)\big]_{st}
 =-\frac{1}{N_c} \sum_{\vect{k}}\sum_{\mu,\nu}\frac{a_\mu^s(\vect{k})a_\mu^{t*}(\vect{k})a_\nu^s(\vect{k+q})a_\nu^{t*}(\vect{k+q})}{i\omega+E_\nu(\vect{k+q})-E_\mu(\vect{k})}\big[n_F(E_\nu(\vect{k+q}))-n_F(E_\mu(\vect{k}))\big] \, ,
\label{chi}
\end{equation} 
where $\mu$, $\nu$ are band indices,  $a_\mu^s(\vect{k})$ and $E_\mu(\vect{k})$ are the $\mu$-th eigenvalue and eigenvector of
the noninteracting Hamiltonian, respectively, and $n_F$ is the Fermi-Dirac distribution function.

The Coulomb interaction can then be included at the mean-field level and one arrives at a so-called
``RPA'' (or ``Stoner'', see, e.g.\ Refs.~\cite{Fazekas99,Pavarini13}) formula for the interacting susceptibility
\begin{equation}
\chi(\vect{q},i\omega)= \frac{\chi_0(\vect{q},i\omega)}{\mathbb{I}-\chi_0(\vect{q},i\omega)\,U} \, ,
\label{eq:ChiStoner}
\end{equation}
where in the paramagnetic state we can use $\chi_0$ according to Eq.~(\ref{chi}). According to
Eq.~(\ref{eq:ChiStoner}), the static susceptibility $\chi(\vect{q},i\omega=0)$ diverges 
whenever $U$ equals one of the eigenvalues of the tensor $\chi_0(\vect{q},i\omega=0)^{-1}$. 
One can use this identity to determine the mean-field critical $U_c$, and indeed, the critical $U_c$ of an infinite graphene
sheet was originally determined in this manner \cite{Sorella1992}. The value of $\vect{q}$ and the corresponding
eigenvector yield information about the expected magnetic state for $U>U_c$.

The tensor of Eq.~(\ref{chi}) is symmetric, but computing all $N_c^2$ entries for a fixed $\vect{q}$ is time-consuming since
each of them involves a sum over reciprocal space and two sums over all energy levels.
In order to keep the CPU time manageable, we have
limited the sum $\sum_{\mu,\nu}$ to states that are close to the Fermi energy.
The latter approximation is physically justified since the ground-state ordering should be dominated by the
quasi-flat bands close to the Fermi energy. A similar approximation to the Matsubara sums has also
been used in Ref.~\cite{Klebl19} except that we use here a more radical sharp cutoff. Nevertheless,
we have checked that taking the $50$ to $100$ states closest to the Fermi energy into account
is sufficient to yield no visible truncation effects; we used $200$ states to be on the safe side.
Since we use a finite grid for the integration over the moir\'e Brillouin zone,
the sum Eq.~(\ref{chi}) consists strictly speaking of a finite number of poles.
In order to smooth these out, we introduce a broadening parameter and evaluate
${\rm Re} \chi_0(\vect{q},i\omega=i\eta)$ such that we obtain a Lorentzian
broadening of width $\eta$ at $i\omega=0$. Apart from the truncation in energy space,
the momentum grid, and the broadening parameter $\eta$, the result for $\chi_0(\vect{q},i\omega=0)$
also depends on temperature $T$.
$T=10^{-8}\,t$ seems to be sufficiently low to ensure ground-state physics.
However, there is a delicate balance between broadening parameter $\eta$ and the grid
in reciprocal space. If $\eta$ is too large, it will smear out any peaks and thus reduce
the values of $\chi(\vect{q},i\omega=0)$. On the other hand, for a too small value of $\eta$, the momentum
discretization will become visible. We found that the combination $\eta=5\cdot10^{-5}\,t$ and a
uniform $9\times 9$ grid of points $(k_x,k_y)$ yield a good compromise such that we will present results
for these parameters here.

%%%%%%%%%%%%%%%%%%%%%%%
%%%%%%%%%%%%%%%%%%%%%%%
\begin{figure}[tb!]
\centering
\includegraphics[width=0.46\linewidth]{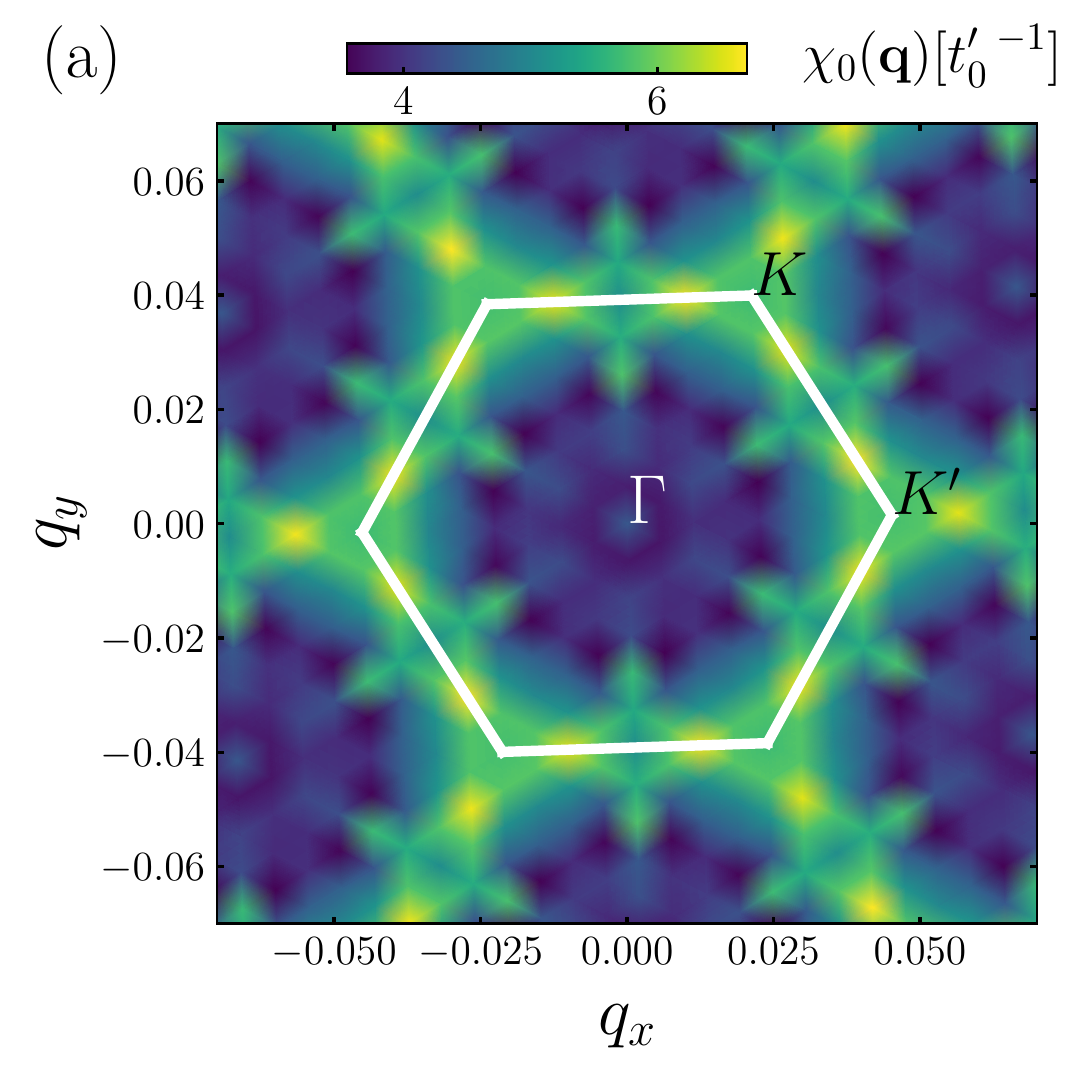}\hfill%
\includegraphics[width=0.48\linewidth]{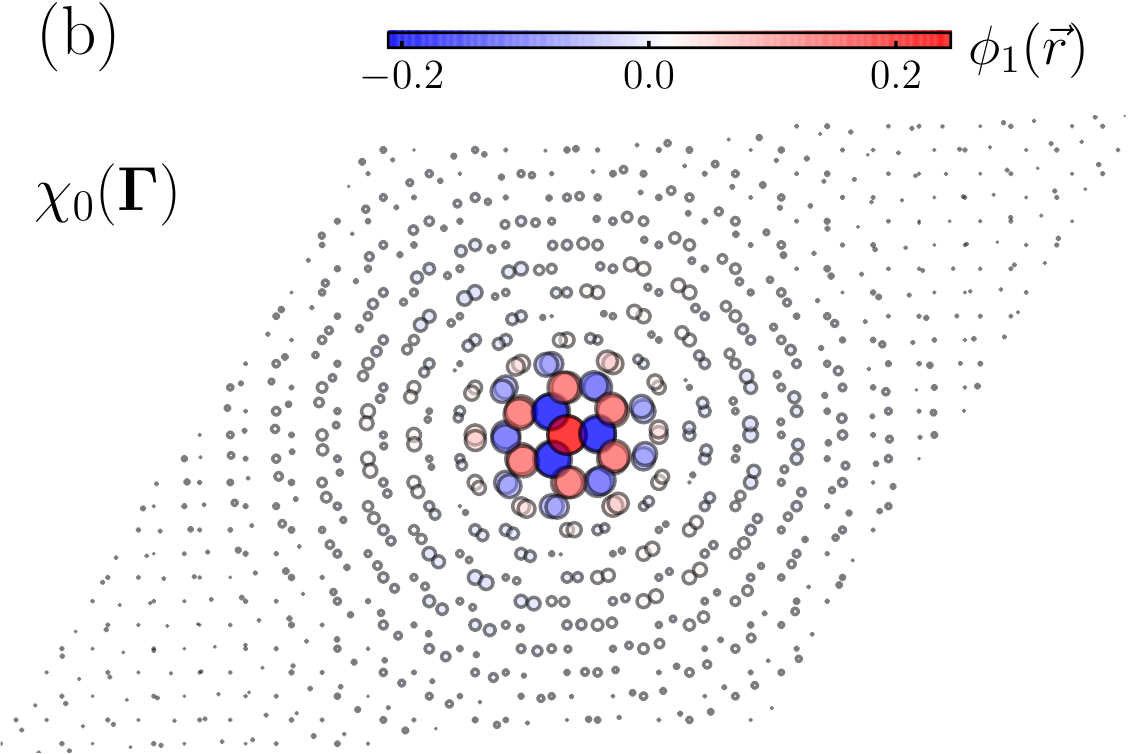}\\
\includegraphics[width=0.48\linewidth]{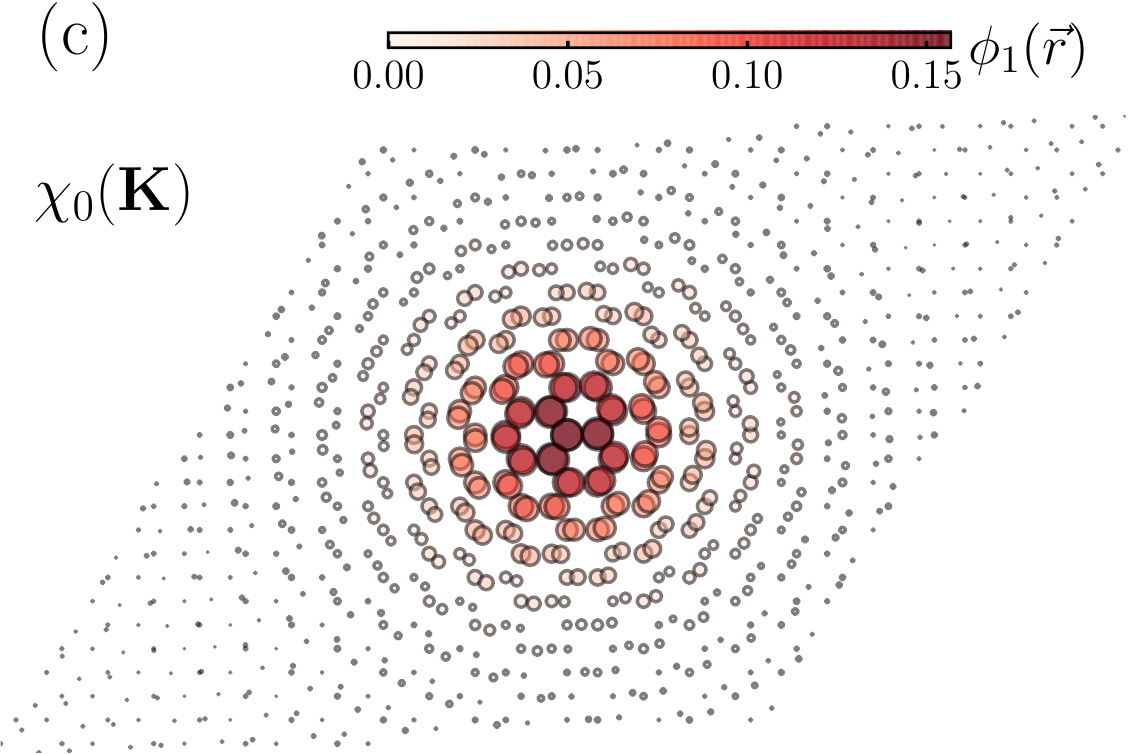}\hfill%
\includegraphics[width=0.48\linewidth]{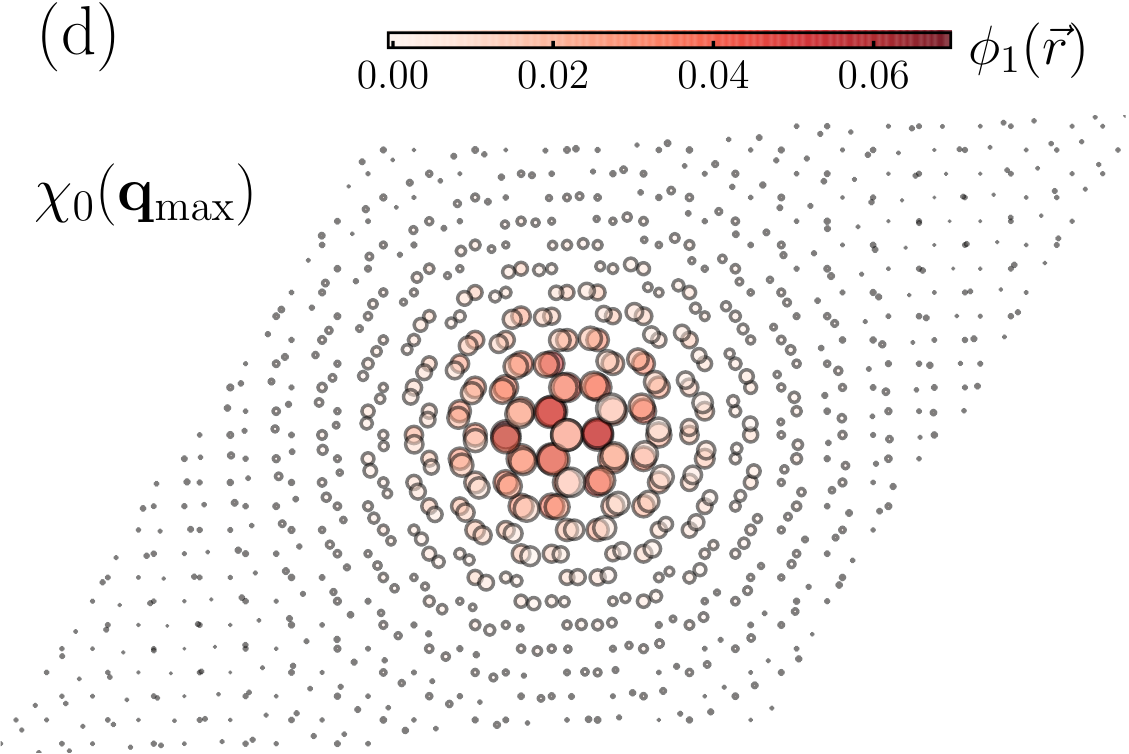}
\caption{(a) Distribution of the largest eigenvalue $\chi_0(\vect{q},0)$
of the susceptibility tensor for the rescaled system
with $\theta_{\rm eff}=1.08^\circ$.
The white hexagon denotes the first Brillouin zone.
Panels (b), (c), and (d) show the spatial profile of the largest eigenvector
of the static susceptibility tensor for
$\vect{q} = \Gamma$, K, and $\vect{q}_{\rm max}$,
respectively. We used $\eta=5\cdot10^{-5}\,t$,
a uniform $9 \times 9$ grid to evaluate the sum $\sum_{\vect{k}}$,
and a total of $200$ states around the Fermi level for each $\mu$ and $\nu$.
}
\label{fig_chi}
\end{figure}
%%%%%%%%%%%%%%%%%%%%%%%
%%%%%%%%%%%%%%%%%%%%%%%

Figure~\ref{fig_chi}(a) shows
the distribution of the leading eigenvalue of the static susceptibility tensor $\chi_0(\vect{q},i\omega=0)$ in the moir\'e Brillouin zone.
In contrast to single layers and AA-stacked bilayer graphene that prefer
a single type of ordering at $\vect{q}=\vect{0}$, in the present case
the maximal eigenvalue of $\chi_0(\vect{q},i\omega=0)$ 
is rather flat in reciprocal space. The global maximum is neither 
at $\vect{q} = \Gamma$ nor at the two symmetry-related
points K and K$'$, but rather at another point $\vect{q}_{\rm max}$ at
the boundary of the first Brillouin zone.
The values are $\max \chi_0(\vect{q},i\omega=0) =
 4.35378/t_0'$, % 4.353779247714188
$4.99189/t_0'$, % 4.991892465834578
and
$6.61892/t_0'$, % 6.618919
for $\vect{q} = \Gamma$, K, and $\vect{q}_{\rm max}$,
respectively.
According to the discussion around Eq.~(\ref{eq:ChiStoner}), this predicts a critical value
$U_c = 0.229686\,t_0'$ % 0.22968550840629273
for a $\vect{q} = \vect{0}$ state and globally
$U_c = 0.151082\,t_0'$, % 0.1510820724653074
but for a state with a spatial modulation with a wave vector
$\vect{q}_{\rm max}$ over moir\'e cells.

Panels (b)--(d) of Fig.~\ref{fig_chi} show the corresponding eigenvectors 
of the susceptibility tensor. At the $\Gamma$ point (panel (b)), one 
observes a staggered sign change between nearest neighbors with the maxima 
located in the AA region. This corresponds to the periodic 
antiferromagnetic state that we have investigated in the main text. The 
analogous result for the eigenvector at the K (K$'$) point is shown in 
Fig.~\ref{fig_chi}(c). Here we find a ferromagnetic solution in each 
moir\'e cell with the maximum again in the AA region, but the 
corresponding value of $\vect{q}$ implies that the corresponding state 
should be accompanied by a tripling of the unit cell in real space. 
Finally, Fig.~\ref{fig_chi}(d) shows the eigenvector at $\vect{q}_{\rm 
max}$. The local structure inside a moir\'e cell is still ferromagnetic, 
but exhibits a stronger internal modulation. Furthermore, the 
corresponding MFT solution should be modulated with a wave vector 
$\vect{q}_{\rm max}$ in real space. Examination of further values of 
$\vect{q}$ reveals an antiferromagnetic internal structure close to the 
$\Gamma$ point while the ferromagnetic internal arrangement is predominant 
in other regions of the Brillouin zone.

\end{appendix}
%%%%%

\bibliography{TGB_Ref}

\begin{thebibliography}{10}
\providecommand{\url}[1]{\texttt{#1}}
\providecommand{\urlprefix}{URL }
\expandafter\ifx\csname urlstyle\endcsname\relax
  \providecommand{\doi}[1]{doi:\discretionary{}{}{}#1}\else
  \providecommand{\doi}{doi:\discretionary{}{}{}\begingroup
  \urlstyle{rm}\Url}\fi
\providecommand{\eprint}[2][]{\url{#2}}

\bibitem{Novoselov04}
K.~S. Novoselov, A.~K. Geim, S.~V. Morozov, D.~Jiang, Y.~Zhang, S.~V. Dubonos,
  I.~V. Grigorieva and A.~A. Firsov,
\newblock \emph{Electric field effect in atomically thin carbon films},
\newblock Science \textbf{306}, 666 (2004),
\newblock \doi{10.1126/science.1102896}.

\bibitem{Novoselov2012}
K.~S. Novoselov, V.~I. Fal'ko, L.~Colombo, P.~R. Gellert, M.~G. Schwab and
  K.~Kim,
\newblock \emph{A roadmap for graphene},
\newblock Nature \textbf{490}, 192 (2012),
\newblock \doi{10.1038/nature11458}.

\bibitem{Novoselov16}
K.~S. Novoselov, A.~Mishchenko, A.~Carvalho and A.~H. Castro~Neto,
\newblock \emph{{2D} materials and van der {W}aals heterostructures},
\newblock Science \textbf{353}, aac9439 (2016),
\newblock \doi{10.1126/science.aac9439}.

\bibitem{Avsar20}
A.~Avsar, H.~Ochoa, F.~Guinea, B.~\"Ozyilmaz, B.~J. van Wees and I.~J.
  Vera-Marun,
\newblock \emph{{\it Colloquium}: Spintronics in graphene and other
  two-dimensional materials},
\newblock Rev. Mod. Phys. \textbf{92}, 021003 (2020),
\newblock \doi{10.1103/RevModPhys.92.021003}.

\bibitem{Yazyev2010}
O.~V. Yazyev,
\newblock \emph{Emergence of magnetism in graphene materials and
  nanostructures},
\newblock Rep. Prog. Phys. \textbf{73}, 056501 (2010),
\newblock \doi{10.1088/0034-4885/73/5/056501}.

\bibitem{Fatemi18}
Y.~Cao, V.~Fatemi, S.~Fang, K.~Watanabe, T.~Taniguchi, E.~Kaxiras and
  P.~Jarillo-Herrero,
\newblock \emph{Unconventional superconductivity in magic-angle graphene
  superlattices},
\newblock Nature \textbf{556}, 43 (2018),
\newblock \doi{10.1038/nature26160}.

\bibitem{Cao18}
Y.~Cao, V.~Fatemi, A.~Demir, S.~Fang, S.~L. Tomarken, J.~Y. Luo, J.~D.
  Sanchez-Yamagishi, K.~Watanabe, T.~Taniguchi, E.~Kaxiras, R.~C. Ashoori and
  P.~Jarillo-Herrero,
\newblock \emph{Correlated insulator behaviour at half-filling in magic-angle
  graphene superlattices},
\newblock Nature \textbf{556}, 80 (2018),
\newblock \doi{10.1038/nature26154}.

\bibitem{andrei2020graphene}
E.~Y. Andrei and A.~H. MacDonald,
\newblock \emph{Graphene bilayers with a twist},
\newblock Nature Materials \textbf{19}, 1265 (2020),
\newblock \doi{10.1038/s41563-020-00840-0}.

\bibitem{tbgI}
B.~A. Bernevig, Z.-D. Song, N.~Regnault and B.~Lian,
\newblock \emph{Twisted bilayer graphene. {I}. {M}atrix elements,
  approximations, perturbation theory, and a
  $k\ifmmode\cdot\else\textperiodcentered\fi{}p$ two-band model},
\newblock Phys. Rev. B \textbf{103}, 205411 (2021),
\newblock \doi{10.1103/PhysRevB.103.205411}.

\bibitem{tbgII}
Z.-D. Song, B.~Lian, N.~Regnault and B.~A. Bernevig,
\newblock \emph{Twisted bilayer graphene. {II}. {S}table symmetry anomaly},
\newblock Phys. Rev. B \textbf{103}, 205412 (2021),
\newblock \doi{10.1103/PhysRevB.103.205412}.

\bibitem{tbgIII}
B.~A. Bernevig, Z.-D. Song, N.~Regnault and B.~Lian,
\newblock \emph{Twisted bilayer graphene. {III}. {I}nteracting {H}amiltonian
  and exact symmetries},
\newblock Phys. Rev. B \textbf{103}, 205413 (2021),
\newblock \doi{10.1103/PhysRevB.103.205413}.

\bibitem{tbgIV}
B.~Lian, Z.-D. Song, N.~Regnault, D.~K. Efetov, A.~Yazdani and B.~A. Bernevig,
\newblock \emph{Twisted bilayer graphene. {IV}. {E}xact insulator ground states
  and phase diagram},
\newblock Phys. Rev. B \textbf{103}, 205414 (2021),
\newblock \doi{10.1103/PhysRevB.103.205414}.

\bibitem{tbgV}
B.~A. Bernevig, B.~Lian, A.~Cowsik, F.~Xie, N.~Regnault and Z.-D. Song,
\newblock \emph{Twisted bilayer graphene. {V}. {E}xact analytic many-body
  excitations in {C}oulomb {H}amiltonians: {C}harge gap, {G}oldstone modes, and
  absence of {C}ooper pairing},
\newblock Phys. Rev. B \textbf{103}, 205415 (2021),
\newblock \doi{10.1103/PhysRevB.103.205415}.

\bibitem{tbgVI}
F.~Xie, A.~Cowsik, Z.-D. Song, B.~Lian, B.~A. Bernevig and N.~Regnault,
\newblock \emph{Twisted bilayer graphene. {VI}. {A}n exact diagonalization
  study at nonzero integer filling},
\newblock Phys. Rev. B \textbf{103}, 205416 (2021),
\newblock \doi{10.1103/PhysRevB.103.205416}.

\bibitem{Senthil18}
H.~C. Po, L.~Zou, A.~Vishwanath and T.~Senthil,
\newblock \emph{Origin of {M}ott insulating behavior and superconductivity in
  twisted bilayer graphene},
\newblock Phys. Rev. X \textbf{8}, 031089 (2018),
\newblock \doi{10.1103/PhysRevX.8.031089}.

\bibitem{Kuroki18}
M.~Ochi, M.~Koshino and K.~Kuroki,
\newblock \emph{Possible correlated insulating states in magic-angle twisted
  bilayer graphene under strongly competing interactions},
\newblock Phys. Rev. B \textbf{98}, 081102(R) (2018),
\newblock \doi{10.1103/PhysRevB.98.081102}.

\bibitem{Pizarro19}
J.~M. Pizarro, M.~J. Calder{\'{o}}n and E.~Bascones,
\newblock \emph{The nature of correlations in the insulating states of twisted
  bilayer graphene},
\newblock J. Phys. Commun. \textbf{3}, 155415 (2019),
\newblock \doi{10.1088/2399-6528/ab0fa9}.

\bibitem{Roy19}
B.~Roy and V.~Juri\v{c},
\newblock \emph{Unconventional superconductivity in nearly flat bands in
  twisted bilayer graphene},
\newblock Phys. Rev. B \textbf{99}, 121407(R) (2019),
\newblock \doi{10.1103/PhysRevB.99.121407}.

\bibitem{MacDonald18}
M.~Xie and A.~H. MacDonald,
\newblock \emph{Nature of the correlated insulator states in twisted bilayer
  graphene},
\newblock Phys. Rev. Lett. \textbf{124}, 097601 (2020),
\newblock \doi{10.1103/PhysRevLett.124.097601}.

\bibitem{Zhang20}
Y.~Zhang, K.~Jiang, Z.~Wang and F.~Zhang,
\newblock \emph{Correlated insulating phases of twisted bilayer graphene at
  commensurate filling fractions: A {H}artree-{F}ock study},
\newblock Phys. Rev. B \textbf{102}, 035136 (2020),
\newblock \doi{10.1103/PhysRevB.102.035136}.

\bibitem{Cea20}
T.~Cea and F.~Guinea,
\newblock \emph{Band structure and insulating states driven by {C}oulomb
  interaction in twisted bilayer graphene},
\newblock Phys. Rev. B \textbf{102}, 045107 (2020),
\newblock \doi{10.1103/PhysRevB.102.045107}.

\bibitem{Klebl20}
L.~Klebl, D.~M. Kennes and C.~Honerkamp,
\newblock \emph{Functional renormalization group for a large moir\'e unit
  cell},
\newblock Phys. Rev. B \textbf{102}, 085109 (2020),
\newblock \doi{10.1103/PhysRevB.102.085109}.

\bibitem{Laksono19}
X.~Gu, C.~Chen, J.~N. Leaw, E.~Laksono, V.~M. Pereira, G.~Vignale and S.~Adam,
\newblock \emph{Antiferromagnetism and chiral $d$-wave superconductivity from
  an effective $t-{J}-{D}$ model for twisted bilayer graphene},
\newblock Phys. Rev. B \textbf{101}, 180506 (2020),
\newblock \doi{10.1103/PhysRevB.101.180506}.

\bibitem{Bultinck20}
N.~Bultinck, E.~Khalaf, S.~Liu, S.~Chatterjee, A.~Vishwanath and M.~P. Zaletel,
\newblock \emph{Ground state and hidden symmetry of magic-angle graphene at
  even integer filling},
\newblock Phys. Rev. X \textbf{10}, 031034 (2020),
\newblock \doi{10.1103/PhysRevX.10.031034}.

\bibitem{liu2019theories}
J.~Liu and X.~Dai,
\newblock \emph{Theories for the correlated insulating states and quantum
  anomalous {H}all effect phenomena in twisted bilayer graphene},
\newblock Phys. Rev. B \textbf{103}, 035427 (2021),
\newblock \doi{10.1103/PhysRevB.103.035427}.

\bibitem{Fazekas99}
P.~Fazekas,
\newblock \emph{Lecture Notes on Electron Correlation and Magnetism},
\newblock World Scientifc, Singapore,
\newblock \doi{10.1142/2945} (1999).

\bibitem{LopesdosSantos07}
J.~M.~B. Lopes~dos Santos, N.~M.~R. Peres and A.~H. Castro~Neto,
\newblock \emph{Graphene bilayer with a twist: Electronic structure},
\newblock Phys. Rev. Lett. \textbf{99}, 256802 (2007),
\newblock \doi{10.1103/PhysRevLett.99.256802}.

\bibitem{Trambly10}
G.~Trambly~de Laissardi\`ere, D.~Mayou and L.~Magaud,
\newblock \emph{Localization of {D}irac electrons in rotated graphene
  bilayers},
\newblock Nano Letters \textbf{10}, 804 (2010),
\newblock \doi{10.1021/nl902948m}.

\bibitem{SuarezMorell10}
E.~Su\'arez~Morell, J.~D. Correa, P.~Vargas, M.~Pacheco and Z.~Barticevic,
\newblock \emph{Flat bands in slightly twisted bilayer graphene: Tight-binding
  calculations},
\newblock Phys. Rev. B \textbf{82}, 121407(R) (2010),
\newblock \doi{10.1103/PhysRevB.82.121407}.

\bibitem{Bistritzer11}
R.~Bistritzer and A.~H. MacDonald,
\newblock \emph{Moir{\'e} bands in twisted double-layer graphene},
\newblock Proceedings of the National Academy of Sciences \textbf{108}, 12233
  (2011),
\newblock \doi{10.1073/pnas.1108174108}.

\bibitem{Trambly12}
G.~Trambly~de Laissardi\`ere, D.~Mayou and L.~Magaud,
\newblock \emph{Numerical studies of confined states in rotated bilayers of
  graphene},
\newblock Phys. Rev. B \textbf{86}, 125413 (2012),
\newblock \doi{10.1103/PhysRevB.86.125413}.

\bibitem{Sharpe19}
A.~L. Sharpe, E.~J. Fox, A.~W. Barnard, J.~Finney, K.~Watanabe, T.~Taniguchi,
  M.~A. Kastner and D.~Goldhaber-Gordon,
\newblock \emph{Emergent ferromagnetism near three-quarters filling in twisted
  bilayer graphene},
\newblock Science \textbf{365}, 605 (2019),
\newblock \doi{10.1126/science.aaw3780}.

\bibitem{pons2020flatband}
R.~Pons, A.~Mielke and T.~Stauber,
\newblock \emph{Flat-band ferromagnetism in twisted bilayer graphene},
\newblock Phys. Rev. B \textbf{102}, 235101 (2020),
\newblock \doi{10.1103/PhysRevB.102.235101}.

\bibitem{Georges1996}
A.~Georges, G.~Kotliar, W.~Krauth and M.~J. Rozenberg,
\newblock \emph{Dynamical mean-field theory of strongly correlated fermion
  systems and the limit of infinite dimensions},
\newblock Rev. Mod. Phys. \textbf{68}, 13 (1996),
\newblock \doi{10.1103/RevModPhys.68.13}.

\bibitem{Marcin19}
M.~Raczkowski, R.~Peters, T.~T. Ph\`ung, N.~Takemori, F.~F. Assaad, A.~Honecker
  and J.~Vahedi,
\newblock \emph{{H}ubbard model on the honeycomb lattice: From static and
  dynamical mean-field theories to lattice quantum {M}onte {C}arlo
  simulations},
\newblock Phys. Rev. B \textbf{101}, 125103 (2020),
\newblock \doi{10.1103/PhysRevB.101.125103}.

\bibitem{Thu2020}
T.~T. Ph\`ung, R.~Peters, A.~Honecker, G.~Trambly~de Laissardi\`ere and
  J.~Vahedi,
\newblock \emph{Spin-caloritronic transport in hexagonal graphene nanoflakes},
\newblock Phys. Rev. B \textbf{102}, 035160 (2020),
\newblock \doi{10.1103/PhysRevB.102.035160}.

\bibitem{Klebl19}
L.~Klebl and C.~Honerkamp,
\newblock \emph{Inherited and flatband-induced ordering in twisted graphene
  bilayers},
\newblock Phys. Rev. B \textbf{100}, 155145 (2019),
\newblock \doi{10.1103/PhysRevB.100.155145}.

\bibitem{Santos12}
J.~M.~B. Lopes~dos Santos, N.~M.~R. Peres and A.~H. Castro~Neto,
\newblock \emph{Continuum model of the twisted graphene bilayer},
\newblock Phys. Rev. B \textbf{86}, 155449 (2012),
\newblock \doi{10.1103/PhysRevB.86.155449}.

\bibitem{Moon13}
P.~Moon and M.~Koshino,
\newblock \emph{Optical absorption in twisted bilayer graphene},
\newblock Phys. Rev. B \textbf{87}, 205404 (2013),
\newblock \doi{10.1103/PhysRevB.87.205404}.

\bibitem{Luis17}
L.~A. Gonzalez-Arraga, J.~L. Lado, F.~Guinea and P.~San-Jose,
\newblock \emph{Electrically controllable magnetism in twisted bilayer
  graphene},
\newblock Phys. Rev. Lett. \textbf{119}, 107201 (2017),
\newblock \doi{10.1103/PhysRevLett.119.107201}.

\bibitem{CastroNeto2009}
A.~H. Castro~Neto, F.~Guinea, N.~M.~R. Peres, K.~S. Novoselov and A.~K. Geim,
\newblock \emph{The electronic properties of graphene},
\newblock Rev. Mod. Phys. \textbf{81}, 109 (2009),
\newblock \doi{10.1103/RevModPhys.81.109}.

\bibitem{Feldner2010}
H.~Feldner, Z.~Y. Meng, A.~Honecker, D.~Cabra, S.~Wessel and F.~F. Assaad,
\newblock \emph{Magnetism of finite graphene samples: Mean-field theory
  compared with exact diagonalization and quantum {M}onte {C}arlo simulations},
\newblock Phys. Rev. B \textbf{81}, 115416 (2010),
\newblock \doi{10.1103/PhysRevB.81.115416};
%
%\bibitem{FeldnerE}
%H.~Feldner, Z.~Y. Meng, A.~Honecker, D.~Cabra, S.~Wessel and F.~F. Assaad,
\newblock \emph{Erratum:}
\newblock Phys. Rev. B \textbf{101}, 049909(E) (2020),
\newblock \doi{10.1103/PhysRevB.101.049909}.

\bibitem{Feldner2011}
H.~Feldner, Z.~Y. Meng, T.~C. Lang, F.~F. Assaad, S.~Wessel and A.~Honecker,
\newblock \emph{Dynamical signatures of edge-state magnetism on graphene
  nanoribbons},
\newblock Phys. Rev. Lett. \textbf{106}, 226401 (2011),
\newblock \doi{10.1103/PhysRevLett.106.226401}.

\bibitem{Sorella1992}
S.~Sorella and E.~Tosatti,
\newblock \emph{Semi-metal-insulator transition of the {H}ubbard model in the
  honeycomb lattice},
\newblock Europhys. Lett. \textbf{19}, 699 (1992),
\newblock \doi{10.1209/0295-5075/19/8/007}.

\bibitem{Sorella2012}
S.~Sorella, Y.~Otsuka and S.~Yunoki,
\newblock \emph{Absence of a spin liquid phase in the {H}ubbard model on the
  honeycomb lattice},
\newblock Sci. Rep. \textbf{2}, 992 (2012),
\newblock \doi{10.1038/srep00992}.

\bibitem{Hassan2013}
S.~R. Hassan and D.~S\'en\'echal,
\newblock \emph{Absence of spin liquid in nonfrustrated correlated systems},
\newblock Phys. Rev. Lett. \textbf{110}, 096402 (2013),
\newblock \doi{10.1103/PhysRevLett.110.096402}.

\bibitem{Assaad2013}
F.~F. Assaad and I.~F. Herbut,
\newblock \emph{Pinning the order: The nature of quantum criticality in the
  {H}ubbard model on honeycomb lattice},
\newblock Phys. Rev. X \textbf{3}, 031010 (2013),
\newblock \doi{10.1103/PhysRevX.3.031010}.

\bibitem{Hirschmeier2018}
D.~Hirschmeier, H.~Hafermann and A.~I. Lichtenstein,
\newblock \emph{Multiband dual fermion approach to quantum criticality in the
  {H}ubbard honeycomb lattice},
\newblock Phys. Rev. B \textbf{97}, 115150 (2018),
\newblock \doi{10.1103/PhysRevB.97.115150}.

\bibitem{Wilson1975}
K.~G. Wilson,
\newblock \emph{The renormalization group: {C}ritical phenomena and the {K}ondo
  problem},
\newblock Rev. Mod. Phys. \textbf{47}, 773 (1975),
\newblock \doi{10.1103/RevModPhys.47.773}.

\bibitem{Krishna1980}
H.~R. Krishna-murthy, J.~W. Wilkins and K.~G. Wilson,
\newblock \emph{Renormalization-group approach to the {A}nderson model of
  dilute magnetic alloys. {I}. {S}tatic properties for the symmetric case},
\newblock Phys. Rev. B \textbf{21}, 1003 (1980),
\newblock \doi{10.1103/PhysRevB.21.1003}.

\bibitem{Bulla2008}
R.~Bulla, T.~A. Costi and T.~Pruschke,
\newblock \emph{Numerical renormalization group method for quantum impurity
  systems},
\newblock Rev. Mod. Phys. \textbf{80}, 395 (2008),
\newblock \doi{10.1103/RevModPhys.80.395}.

\bibitem{Robert2014}
R.~Peters and N.~Kawakami,
\newblock \emph{Spin density waves in the {H}ubbard model: A {DMFT} approach},
\newblock Phys. Rev. B \textbf{89}, 155134 (2014),
\newblock \doi{10.1103/PhysRevB.89.155134}.

\bibitem{Robert2015}
R.~Peters and N.~Kawakami,
\newblock \emph{Large and small {F}ermi-surface spin density waves in the
  {K}ondo lattice model},
\newblock Phys. Rev. B \textbf{92}, 075103 (2015),
\newblock \doi{10.1103/PhysRevB.92.075103}.

\bibitem{gonzalez2021magnetic}
J.~Gonz\'alez and T.~Stauber,
\newblock \emph{Magnetic phases from competing {H}ubbard and extended {C}oulomb
  interactions in twisted bilayer graphene},
\newblock Phys. Rev. B \textbf{104}, 115110 (2021),
\newblock \doi{10.1103/PhysRevB.104.115110}.

\bibitem{Koshino18}
M.~Koshino, N.~F.~Q. Yuan, T.~Koretsune, M.~Ochi, K.~Kuroki and L.~Fu,
\newblock \emph{Maximally localized {W}annier orbitals and the extended
  {H}ubbard model for twisted bilayer graphene},
\newblock Phys. Rev. X \textbf{8}, 031087 (2018),
\newblock \doi{10.1103/PhysRevX.8.031087}.

\bibitem{Kang18}
J.~Kang and O.~Vafek,
\newblock \emph{Symmetry, maximally localized {W}annier states, and a
  low-energy model for twisted bilayer graphene narrow bands},
\newblock Phys. Rev. X \textbf{8}, 031088 (2018),
\newblock \doi{10.1103/PhysRevX.8.031088}.

\bibitem{Vafek19}
J.~Kang and O.~Vafek,
\newblock \emph{Strong coupling phases of partially filled twisted bilayer
  graphene narrow bands},
\newblock Phys. Rev. Lett. \textbf{122}, 246401 (2019),
\newblock \doi{10.1103/PhysRevLett.122.246401}.

\bibitem{davydov2020}
A.~Davydov, K.~Choo, M.~H. Fischer and T.~Neupert,
\newblock \emph{Four- and twelve-band low-energy symmetric {H}amiltonians and
  {H}ubbard parameters for twisted bilayer graphene using ab-initio input}
  (2020), \eprint{https://arxiv.org/abs/2012.12942}.

\bibitem{klebl2020}
L.~Klebl, Z.~A.~H. Goodwin, A.~A. Mostofi, D.~M. Kennes and J.~Lischner,
\newblock \emph{Importance of long-ranged electron-electron interactions for
  the magnetic phase diagram of twisted bilayer graphene},
\newblock Phys. Rev. B \textbf{103}, 195127 (2021),
\newblock \doi{10.1103/PhysRevB.103.195127}.

\bibitem{Xie19}
Y.~Xie, B.~Lian, B.~J{\"a}ck, X.~Liu, C.~L. Chiu, K.~Watanabe, T.~Taniguchi,
  B.~A. Bernevig and A.~Yazdani,
\newblock \emph{Spectroscopic signatures of many-body correlations in
  magic-angle twisted bilayer graphene},
\newblock Nature \textbf{572}, 101 (2019),
\newblock \doi{10.1038/s41586-019-1422-x}.

\bibitem{Kerelsky2019}
A.~Kerelsky, L.~J. McGilly, D.~M. Kennes, L.~Xian, M.~Yankowitz, S.~Chen,
  K.~Watanabe, T.~Taniguchi, J.~Hone, C.~Dean, A.~Rubio and A.~N. Pasupathy,
\newblock \emph{Maximized electron interactions at the magic angle in twisted
  bilayer graphene},
\newblock Nature \textbf{572}, 95 (2019),
\newblock \doi{10.1038/s41586-019-1431-9}.

\bibitem{Choi2019}
Y.~Choi, J.~Kemmer, Y.~Peng, A.~Thomson, H.~Arora, R.~Polski, Y.~Zhang, H.~Ren,
  J.~Alicea, G.~Refael, F.~von Oppen, K.~Watanabe \emph{et~al.},
\newblock \emph{Electronic correlations in twisted bilayer graphene near the
  magic angle},
\newblock Nature Physics \textbf{15}, 1174 (2019),
\newblock \doi{10.1038/s41567-019-0606-5}.

\bibitem{Jiang2019}
Y.~Jiang, X.~Lai, K.~Watanabe, T.~Taniguchi, K.~Haule, J.~Mao and E.~Y. Andrei,
\newblock \emph{Charge order and broken rotational symmetry in magic-angle
  twisted bilayer graphene},
\newblock Nature \textbf{573}, 91 (2019),
\newblock \doi{10.1038/s41586-019-1460-4}.

\bibitem{Huder2018}
L.~Huder, A.~Artaud, T.~Le~Quang, G.~Trambly~de Laissardi\`ere, A.~G.~M.
  Jansen, G.~Lapertot, C.~Chapelier and V.~T. Renard,
\newblock \emph{Electronic spectrum of twisted graphene layers under
  heterostrain},
\newblock Phys. Rev. Lett. \textbf{120}, 156405 (2018),
\newblock \doi{10.1103/PhysRevLett.120.156405}.

\bibitem{Mesple2020}
F.~Mesple, A.~Missaoui, T.~Cea, L.~Huder, F.~Guinea, G.~Trambly~de
  Laissardi\`ere, C.~Chapelier and V.~T. Renard,
\newblock \emph{Heterostrain determines flat bands in magic-angle twisted
  graphene layers},
\newblock Phys. Rev. Lett. \textbf{127}, 126405 (2021),
\newblock \doi{10.1103/PhysRevLett.127.126405}.

\bibitem{Liao21}
Y.~Da~Liao, J.~Kang, C.~N. Brei\o{}, X.~Y. Xu, H.-Q. Wu, B.~M. Andersen, R.~M.
  Fernandes and Z.~Y. Meng,
\newblock \emph{Correlation-induced insulating topological phases at charge
  neutrality in twisted bilayer graphene},
\newblock Phys. Rev. X \textbf{11}, 011014 (2021),
\newblock \doi{10.1103/PhysRevX.11.011014}.

\bibitem{Wilhelm21}
P.~Wilhelm, T.~C. Lang and A.~M. L\"auchli,
\newblock \emph{Interplay of fractional {C}hern insulator and charge density
  wave phases in twisted bilayer graphene},
\newblock Phys. Rev. B \textbf{103}, 125406 (2021),
\newblock \doi{10.1103/PhysRevB.103.125406}.

\bibitem{pahlevanzadeh2021}
B.~Pahlevanzadeh, P.~Sahebsara and D.~S\'en\'echal,
\newblock \emph{{Chiral $p$-wave superconductivity in twisted bilayer graphene
  from dynamical mean field theory}},
\newblock SciPost Phys. \textbf{11}, 17 (2021),
\newblock \doi{10.21468/SciPostPhys.11.1.017}.

\bibitem{potasz2021}
P.~Potasz, M.~Xie and A.~H. MacDonald,
\newblock \emph{Exact diagonalization for magic-angle twisted bilayer
  graphene},
\newblock Phys. Rev. Lett. \textbf{127}, 147203 (2021),
\newblock \doi{10.1103/PhysRevLett.127.147203}.

\bibitem{Maier09}
S.~Graser, T.~A. Maier, P.~J. Hirschfeld and D.~J. Scalapino,
\newblock \emph{Near-degeneracy of several pairing channels in multiorbital
  models for the {F}e pnictides},
\newblock New J. Phys. \textbf{11}, 025016 (2009),
\newblock \doi{10.1088/1367-2630/11/2/025016}.

\bibitem{Liu18}
C.-C. Liu, L.-D. Zhang, W.-Q. Chen and F.~Yang,
\newblock \emph{Chiral spin density wave and $d+id$ superconductivity in the
  magic-angle-twisted bilayer graphene},
\newblock Phys. Rev. Lett. \textbf{121}, 217001 (2018),
\newblock \doi{10.1103/PhysRevLett.121.217001}.

\bibitem{Pavarini13}
E.~Pavarini,
\newblock \emph{Magnetism: Models and Mechanisms}, vol.~3 of \emph{Modeling and
  Simulation}, chap.~3, pp. 3.1--3.44,
\newblock Forschungszentrum J\"ulich Zentralbibliothek, Verlag, J\"ulich,
\newblock ISBN 978-3-89336-884-6 (2013).

\end{thebibliography}

\nolinenumbers

\end{document}